\documentclass[aps,pre,preprint,showpacs,superscriptaddress]{revtex4}
\usepackage{amsbsy}
\usepackage{amssymb}
\usepackage{graphicx}

\newcommand{\vectornorm}[1]{\left|#1\right|}
\newcommand{\sign}{\mathrm{sign}}

\newcommand{\imaginary}[1]{\mathcal{I}m\left(#1\right)}
\def\Xint#1{\mathchoice
   {\XXint\displaystyle\textstyle{#1}}%
   {\XXint\textstyle\scriptstyle{#1}}%
   {\XXint\scriptstyle\scriptscriptstyle{#1}}%
   {\XXint\scriptscriptstyle\scriptscriptstyle{#1}}%
   \!\int}
\def\XXint#1#2#3{{\setbox0=\hbox{$#1{#2#3}{\int}$}
     \vcenter{\hbox{$#2#3$}}\kern-.5\wd0}}

\def\dashint{\Xint-}
\begin{document}

\title{The Sivashinsky equation for corrugated flames in the large-wrinkle limit}

\author{Guy Joulin}
\thanks{Corresponding author}
\email{joulin@lcd.ensma.fr}
\affiliation{Laboratoire de Combustion et de D\'etonique, UPR 9028 du CNRS, ENSMA, 1 rue Cl\'ement
Ader, B.P. 40109, 86961 Futuroscope Cedex, Poitiers, France.}
\author{Bruno Denet}
\email{bruno.denet@irphe.univ-mrs.fr}
\affiliation{Institut de Recherche sur les Ph\'enom\`enes Hors d'Equilibre, UMR 6594 du CNRS,
Technopole de Ch\^ateau Gombert, 49 rue Joliot-Curie, 13384 Marseille Cedex 13, France.}

\date{\today}

\begin{abstract}
Sivashinsky's (1977) nonlinear integro-differential equation for the shape of corrugated
1-dimensional flames is ultimately reducible to a $2N$-body problem, involving the $2N$ complex
poles of the flame slope. Thual, Frisch \& Henon (1985) derived singular linear integral equations
for the pole density in the limit of large steady wrinkles $(N \gg 1)$, which they solved exactly for monocoalesced periodic fronts of highest amplitude of wrinkling and approximately otherwise. Here we solve those analytically for isolated crests, next for monocoalesced then bicoalesced periodic flame patterns, whatever the (large-) amplitudes involved. We compare the
analytically predicted pole densities and flame shapes to numerical results deduced from the
pole-decomposition approach. Good agreement is obtained, even for moderately large $N$s. The results
are extended  to give hints as to the dynamics of supplementary poles. Open problems are evoked.
\end{abstract}

%\keywords{}

\maketitle
\section{Introduction}
Being able to describe the nonlinear development of the Landau-Darrieus \cite{landau,darrieus} (LD)
instability of
premixed-flame fronts is a central topic in combustion theory. As early as 1977 Sivashinsky
\cite{shiva77} showed, in the limit $\mathcal{A} \ll 1$ of small Attwood numbers
based upon the fresh gas ($\rho_u$) or burnt gas ($\rho_b<\rho_u$) densities, $0 < \mathcal{A}\equiv(\rho_u-\rho_b)/(\rho_u+\rho_b) <1$, that the shape $\phi(x,t)$ of a
flat-on-average, spontaneously evolving wrinkled flame is governed by
\begin{equation}
\label{eq:siveq}
\displaystyle
\phi_t + \frac{1}{2} \phi_x^2 = \nu\,\phi_{xx} + I({\phi}) 
\end{equation}
in suitable units. In (\ref{eq:siveq}) the subscripts denote partial derivatives with respect to
time, $t$, and coordinate, $x$, normal to the mean direction of propagation, and the ``viscosity''
$\nu
> 0$ represents a reciprocal Peclet number based upon the actual flame thickness and the wrinkle
wavelength. The linear integral operator $I(\cdot)$ is defined by $I(e^{ikx})=\vectornorm{k}
e^{ikx}$ (whence $I({\phi})$ is the Hilbert transform, $\hat{H}(-\phi_x)$, of $-\phi_x$) and stems
from the LD instability. The growth/decay rate of infinitesimal harmonics is
$\left|k\right|- \nu k^2$, which identifies $1/\nu$ and $\nu$ as neutral wavenumber and minimum growth
time, respectively. The nonlinearity is geometrical, accounting as it does for the cosine, 
$(1+s^2)^{-1/2}\simeq 1-s^2/2+\ldots$, of
the small angle ($\arctan(s) \simeq s+\ldots$) that the local normal to the flame front makes to the mean direction
of propagation, where $s\sim\phi_x\times \mathcal{A}$ is the unscaled
front slope. Originally derived in \cite{shiva77} as a leading order result for $\mathcal{A}
\rightarrow 0^+$, equation (\ref{eq:siveq}) happens to govern the shape of steadily propagating
fronts even when two more terms of the $\mathcal{A}$-expansion are retained \cite{sivclav,kak05}; its
structure then remains valid practically up to $\mathcal{A}=3/4$, {\it i.e.}, $\rho_u=7\rho_b$
\cite{kak05}.

Numerics \cite{michelson77} reveals that ``steady''  solutions of (\ref{eq:siveq}), corresponding to
$\phi(x,t) = -V t + \phi(x)$ are often ultimately reached. When (\ref{eq:siveq}) is integrated with
periodic boundary conditions for ``not-too-small" values of $\nu$, $\nu > 1/25$ say, the ``steady"
pattern has a single crest per $x$-wise interval of $2\pi$ length, where $\phi_{xx}$ is large and
negative; without loss of generality one may assume that one is  located at $x=0$, in which
case $\phi_x = 0$ when $x$ is an integer multiple of $\pi$ (i.e., $x=0$ ($\bmod~\pi$)) and $\phi_{xx}(\pm\pi)\simeq 1/\pi$. If Neumann conditions at
$x=0$ and $x=\pi$ are used instead, still with a moderately small $\nu$, the final
pattern obtained from
numerical (pseudo-spectral) integrations of (\ref{eq:siveq}) may also have an extra crest located at
$x=\pi$ \cite{denet06}, with $\phi_{xx}(\pi)$ large and negative. By the very way they are
obtained as final state of an unsteady process the 2-crested patterns have a finite basin of
attraction, contrary to the case of
periodic boundary conditions \cite{denet06} where the only stable patterns have a single crest per
cell; yet such ``half-channel'' solutions happen to coincide with the restriction to $0\le x \le
\pi$ of properly shifted $2\pi$-periodic ones, for these are symmetric about $x=0$ and $x=\pi$. If
$\nu$ is too small the widest patterns get very sensitive to noise, even
when caused by numerical rounding-off. In \cite{joulin89} the estimate $\mu \ge O(e^{-1/2\nu\kappa}) \equiv \mu_c(\kappa)$ was
obtained for the noise intensity $\mu$ needed to trigger the appearance of extra-cells on top of the main ones
with periodic boundary conditions; the number $\kappa$ in the above exponent is $\phi_{xx}(\pm\pi) \simeq 1/\pi$; since the most rapidly growing noise-induced disturbances (with initial wavenumbers $\left|k\right|\simeq 1/\nu$ \cite{joulin89}) of a nearly parabolic trough undergo an $O(e^{1/2\nu\kappa})$ amplification, they ultimately get visible as subwrinkles of $O(1)$ final amplitude if $\mu \ge \mu_c(\kappa)$.
Having a larger $\phi_{xx} > 0$ at their troughs (see Sec. 7), 2-crested patterns are presumably less sensitive to noise than the single-crest ones associated with the same wavelength, because $\mu_c(\kappa)$ increases dramatically with $\kappa$ when $\nu$ is small. The 
numerical work of Ref. \cite{denet07} also showed that sums $\phi(x_1,x_2,t) =\phi_1(x_1,t)+\phi_2(x_2,t)$ of
orthogonal, 2-crested one-dimensional patterns play a central role in the study of
(\ref{eq:siveq}) generalized to 2-dimensional flames ($x\rightarrow (x_1,x_2), \phi_x^2\rightarrow \left|\boldsymbol{\nabla}\phi\right|^2$, $\phi_{xx} \rightarrow \Delta \phi$, $I(\cdot) \equiv$ multiplication by $(\boldsymbol{k}\cdot\boldsymbol{k})^{1/2}$ in the 2-D Fourier space $\boldsymbol{k}=(k_1,k_2)$) and to rectangular domains in the Cartesian $(x_1,x_2)$ plane. Without noise such sums are exact stable solutions; with random additive forcing they recurrently appear as long-lived transients when Neumann
conditions are adopted.

Further analyses on the stability of solutions of Eq. (\ref{eq:siveq}) and their responses thus seem
warranted, and getting the ``steady" patterns that correspond to wide, hence large,  cells (or small
$\nu$s) is a
prerequisite. The present contribution is intended to do this.

It is organised as follows. Section 2 introduces the pole-decomposition method, the discrete
equations for the pole locations, and the two integral equations that approximate them for large front wrinkles. The latter equations are next solved analytically for isolated crests (Sec. 3) then one-crested
periodic patterns (Section 4), and the prediction compared to numerical results from the
pole-decomposition approach. Sections 5 and 6 compute the flame speed from the density, and take up the dynamics of a few extra-poles, respectively. Section 7 generalizes the above integral equations
to a pair of coupled ones corresponding to 2-crest periodic flames (and ``half-channel'' ones), then
solves them analytically; comparisons with numerics are again presented. We end up with concluding remarks and open problems
(Sec. 8).

\section{Pole-decomposition(s)}
In 1985 Thual, Frisch and Henon \cite{tfh85} (herein referred to as ``TFH") discovered (see also
\cite{lee-chen}) that (\ref{eq:siveq}) possesses solutions $\phi(x,t)$ representing $2\pi$-periodic
flame patterns with slopes $\phi_x$ in the form
\begin{equation}
\label{eq:2.1}
\displaystyle
\phi_x(x,t) = -\nu \sum_{\alpha=-N}^{N} \cot\left(\frac{x-z_{\alpha}}{2}\right),
\end{equation}
in which the complex-valued poles of $\phi_x(x,t)$, $z_{\alpha}(t)$, are involved in conjugate
pairs ($z_{-\alpha}=z^*_{\alpha}$, $\alpha \neq 0$) for $\phi_x(x,t)$ to be real when $x$ is. For
this pole-decomposed expression to solve (\ref{eq:siveq}), the $z'_{\alpha}$s ($\alpha = -N,\ldots,
-1,1, \ldots, N$) must evolve according to the $2N$-body problem
\begin{equation}
\label{eq:2.2}
\displaystyle
\frac{dz_{\alpha}}{dt_{\phantom{\alpha}}} = \nu \mathop{\sum_{\beta=-N}^{N}}_{\beta \neq \alpha}
\cot\left(\frac{z_{\beta}-z_{\alpha}}{2}\right) -i \,\sign(\imaginary{z_{\alpha}}),
\end{equation}
where $\imaginary{\cdot}$ denotes the imaginary parts of ($\cdot$) and the signum function (with
$\sign (0) = 0$) accounts for the LD instability. Once (\ref{eq:2.2}) is solved for the pole
locations, $\phi(x,t)$ is available from (\ref{eq:2.1}) and the wrinkling-induced excess propagation
speed $V =
-\langle\phi_t\rangle >0$ follows from (\ref{eq:siveq}):
\begin{equation}
\label{eq:2.3}
\displaystyle
V = \frac{1}{2} \left\langle\phi_x^2\right\rangle,
\end{equation}
where $\langle\cdot\rangle$ stands for an average along the $x$-coordinate; thus, $V$ simply measures the wrinkling-induced fractional increase in flame arclength, since $\langle(1+s^2)^{1/2}-1\rangle= \langle s^2/2\rangle +\ldots\sim \mathcal{A}^2\times V$. Beside periodic
$\phi(x,t)$s,
(\ref{eq:siveq}) also allows \cite{tfh85} for isolated non periodic wrinkles that have an infinite wavelength, $V=0$, $\cot(z)$ replaced by $1/z$, and 
\begin{equation}
\label{eq:2.4}
\displaystyle
\frac{dz_{\alpha}}{dt_{\phantom{\alpha}}} = \nu \mathop{\sum_{\beta=-N}^{N}}_{\beta \neq \alpha}
\frac{2}{z_{\beta}-z_{\alpha}} -i\,\sign(\imaginary{z_{\alpha}}).
\end{equation}
In the latter situation, the precise value of $\nu > 0$ does not matter since it could be scaled
out, and the integer $N \ge 1$ is arbitrary. As for (\ref{eq:2.1}) (\ref{eq:2.2}), the maximum
allowed value $N_{\mathrm{opt}}(\nu)$ of $N$ in steady configurations increases with $1/\nu > 1$
\cite{tfh85}. As shown by
TFH, steady flames obtained from (\ref{eq:2.2}) or (\ref{eq:2.4}) correspond to poles that
``coalesce" (or align) along parallels to the imaginary axis, as a result of the pairwise pole
interactions that are attractive along the real $x$-axis and repulsive in the normal direction. In
the case of an isolated crest located at $x=0$, the poles ultimately involved in steady
solution are of the form $i B_{\alpha}$, $-N \le \alpha \le N$, $\alpha\neq 0$, with real $B_{\alpha}$s
satisfying coupled discrete equations deduced from (\ref{eq:2.4}):
\begin{equation}\label{eq:2.4b}
 \nu \mathop{\sum_{\beta=-N}^{N}}_{\beta \neq \alpha}
\frac{2}{B_{\alpha}-B_{\beta}} = \sign(B_{\alpha}).
\end{equation}

The authors of Ref. \cite{tfh85} also evidenced that the larger the number $N$ of pole-pairs in such
``vertical" steady alignments, the smoother the involved poles are distributed along the $B$ coordinate,
 with $B_{\alpha+1}-B_{\alpha}$ well smaller than $B_N$. This suggested TFH to replace the
discrete sum in (\ref{eq:2.4b}) (or its analogue deduced from (\ref{eq:2.2})) by an integral over the
continuous variable $B$, with such a continuous measure that $P(B)\,dB$ is the number of poles located between $B$ and $B + dB$; a constructive definition of $P(B)$ is specified in (\ref{eq:3.8b}). In this continuous approximation the steady
versions of (\ref{eq:2.1}) (\ref{eq:2.2}) are amenable to singular Fredholm integral equations, specifically:
\begin{equation}
\label{eq:2.5}
\displaystyle
\dashint\frac{2\nu P(B')}{B-B'}\,dB' = \sign(B)
\end{equation}
in the non-periodic situations (an isolated wrinkle at $x = 0$), and
\begin{equation}
\label{eq:2.6}
\displaystyle
\dashint \nu P(B') \coth\left(\frac{B-B'}{2}\right)\,dB' = \sign(B)
\end{equation}
for the monocoalesced $2\pi$-periodic cases (one single crest per cell, at $x = 0$ ($\bmod~2\pi$)).
In (\ref{eq:2.5}) (\ref{eq:2.6}) $B$ denotes the pole imaginary coordinate, and the Cauchy principal
parts $\dashint \cdot\,dB'$ stem from the condition $\beta \neq \alpha$ on the sums featured in
(\ref{eq:2.2}) (\ref{eq:2.4}). Consistent with their interpretation as pole densities, the $P(B)$s
showing up in (\ref{eq:2.5}) (\ref{eq:2.6}) both are non-negative even functions of their
argument (for $\phi_x$ to be real when $x$ is) and are normalized by
\begin{equation}
\label{eq:2.7}
\int P(B')\,dB' = 2N.
\end{equation}
In (\ref{eq:2.5})-(\ref{eq:2.7}) the integrals extend over the ranges (to be determined as part of
the solutions) where $P(B) \neq 0$. The next sections will solve (\ref{eq:2.5}) (\ref{eq:2.6})
(\ref{eq:2.7}) analytically, starting with the simpler equation (\ref{eq:2.5}). 

\section{Isolated crest}
Because isolated crests have $\phi_x \rightarrow 0$ at $\left|x\right| \rightarrow \infty$,  we
firstly anticipate the existence of some finite $B_{\mathrm{max}} > 0$ such that
$P(\left|B\right|>B_{\mathrm{max}})\equiv 0$ in (\ref{eq:2.5}). We next recall the identity
\begin{equation}
\label{eq:3.1}
\displaystyle
\dashint_{-\pi/2}^{\pi/2}\frac{\cos((2M+1)\Phi') \cos \Phi'}{\sin\Phi - \sin \Phi'}\,d\Phi' = \pi
\sin((2M+1)\Phi)
\end{equation}
that can be deduced, through the change of variable $\Phi\rightarrow\Phi +
\pi/2$, from a similar one appearing in the Prandtl theory of lifting
lines \cite{batchelor,landau-lifschitz}. Identity (\ref{eq:3.1}) allows one to solve such singular
integral equations as Wigner's \cite{mehta} (for the density, $2\nu P$ say, of eigenvalues of large
real random matrices in the Gaussian Orthogonal Ensemble), written here as
\begin{equation}
\label{eq:3.2}
\displaystyle
\dashint \frac{2\nu P(B')}{B-B'}\,dB' = B;
\end{equation}
its solution is the celebrated semi-circle law $2\pi\nu P(B) = \mathrm{max}(B_{\mathrm{max}}\cos
\Phi,0)$,
\cite{mehta}, provided that one sets
\begin{equation}
\label{eq:3.3}
\displaystyle
B = B_{\mathrm{max}} \sin\Phi,\qquad -\frac{\pi}{2} \le \Phi \le \frac{\pi}{2},
\end{equation}
in (\ref{eq:3.2}). Interestingly, the same change of independent variable in (\ref{eq:2.5})
produces
\begin{equation}
\label{eq:3.4}
\displaystyle
\dashint_{-\pi/2}^{\pi/2} \frac{2\nu P(\Phi')\cos\Phi'}{\sin\Phi-\sin\Phi'}\,d\Phi' = \sign(\Phi),
\end{equation}
since $\sign(B) = \sign(\Phi)$ for $\left|\Phi\right| < \pi$. Over the same range (and hence over
the narrower support of $P$, $\left|\Phi\right| \le \pi/2$), the right-hand side of (\ref{eq:3.4})
may be expanded as the Fourier series
\begin{equation}
\label{eq:3.5}
\displaystyle
\sign(\Phi) = \frac{4}{\pi} \sum_{M=0}^{\infty} \frac{1}{2M +1} \sin((2M+1) \Phi),
\end{equation}
consistent with our convention that $\sign(0)=0$. From (\ref{eq:3.1}) the solution to
(\ref{eq:3.4}) can thus be written as a Fourier series of cosines that all vanish at  $\Phi=\pm\pi/2$:
\begin{eqnarray}
2\nu P(\Phi) & = & \frac{4}{\pi^2} \sum_{M=0}^{\infty} \frac{1}{2M +1} \cos((2M+1) \Phi)\label{eq:3.6a}\\
                    & = & \frac{1}{\pi^2}\log\left(\frac{1+\cos \Phi}{1-\cos
\Phi}\right)\label{eq:3.6c}\\
                    & = &
\frac{1}{\pi^2}\log\left(\frac{1+\sqrt{1-B^2/B^2_{\mathrm{max}}}}{1-\sqrt{1-B^2/B^2_{\mathrm{ max}}
} }\right),\label{eq:3.6d}
\end{eqnarray}%
and $P \equiv 0$ for $\left|B\right| > B_{\mathrm{max}}$; to get (\ref{eq:3.6d}) from (\ref{eq:3.6c}), (\ref{eq:3.3}) was
explicitly employed.

\begin{figure}
  \includegraphics[width=\textwidth]{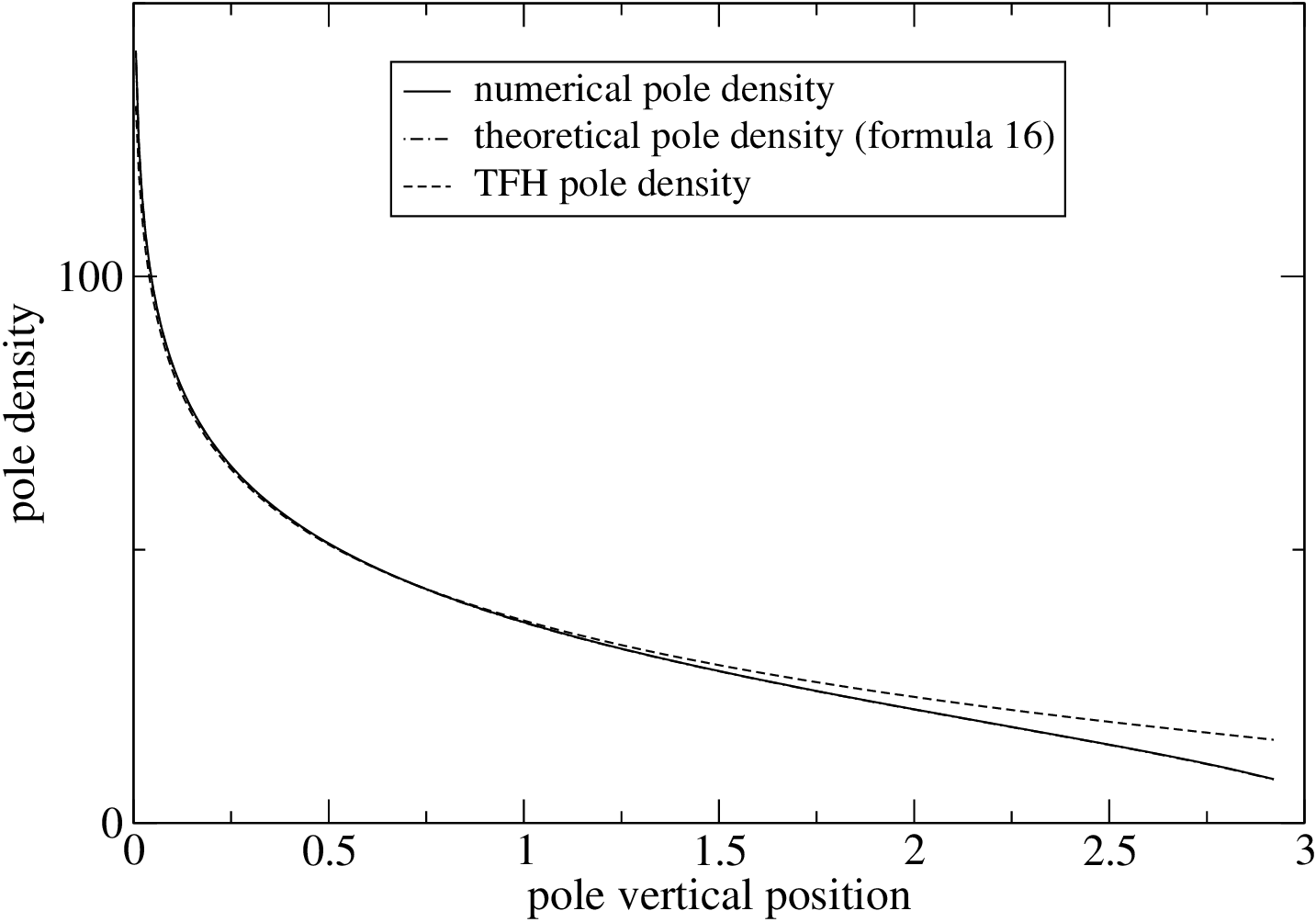}
  \caption{\label{fig:3}Numerical ((\ref{eq:3.8b}), solid line) vs analytical ((\ref{eq:3.6d}),
dot-dashed line) pole densities $P(B)$ for an isolated crest with $1/\nu = 199.5$, $N = 100$. The
dashed line is the TFH fit.}
\end{figure}

\begin{figure}
  \includegraphics[width=\textwidth]{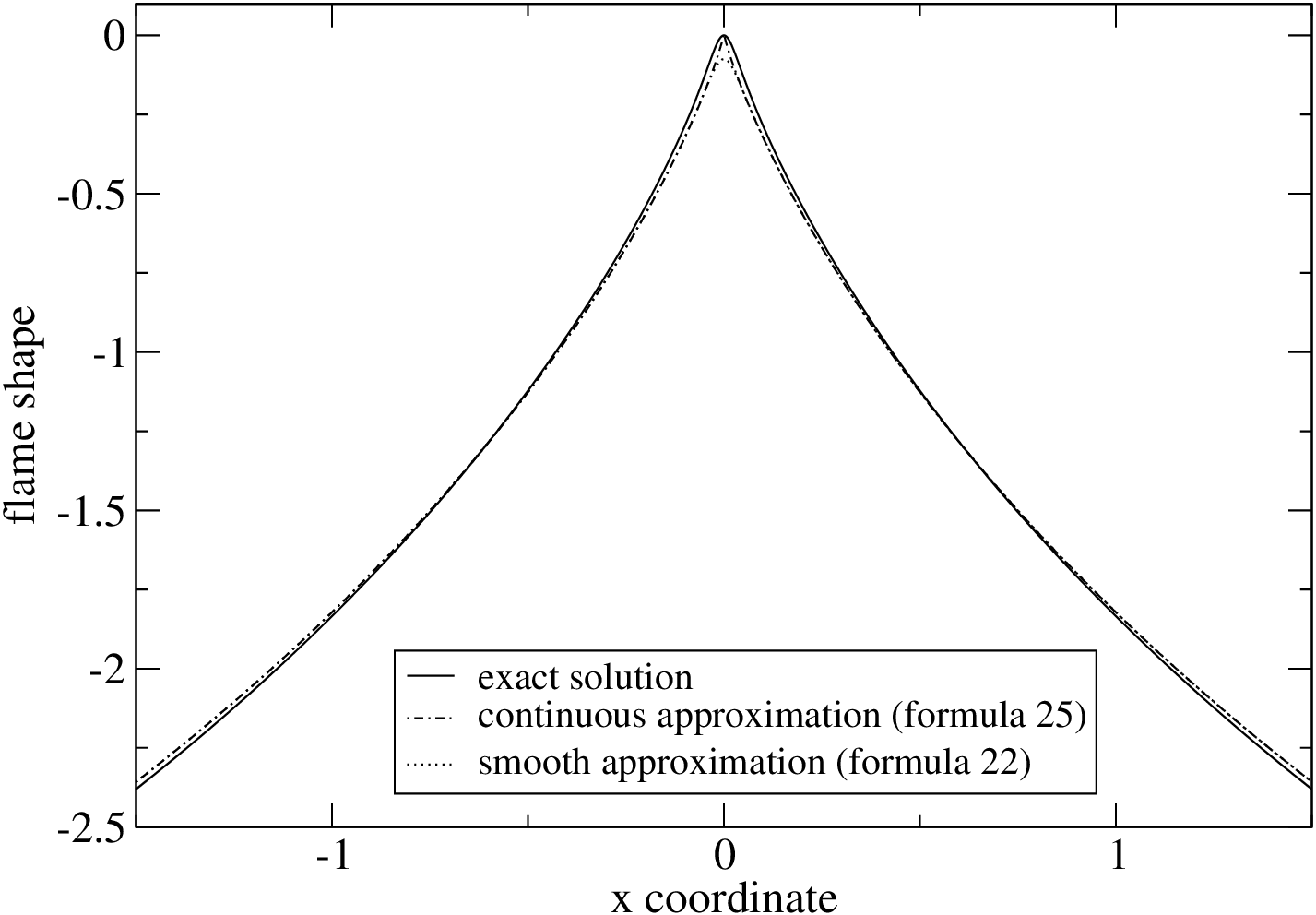}
  \caption{\label{fig:4}Shapes of an isolated crest with $1/\nu =19.5$, $N = 10$: continuous
approximation ((\ref{eq:25}), dot-dashed line), exact (solid line), and smooth approximation from
eq. (\ref{eq:3.10}) (dotted).}
\end{figure}

The cumulative pole distribution $R(B)=\int_0^B P(B')\,dB'$ reads, after integration by parts, as
\begin{equation}
\label{eq:3.7}
\displaystyle
2\nu R(B) = \frac{B_{\mathrm{max}}}{\pi^2}\left(\sin\Phi \log\frac{1+\cos\Phi}{1-\cos\Phi} +
2\Phi\right),
\end{equation}
whereby the renormalization condition $R(B_{\mathrm{max}}) = R(\Phi=\pi/2) = N$ fixes
$B_{\mathrm{max}}$ to be given by
\begin{equation}
\label{eq:3.8}
B_{\mathrm{max}} = 2\pi N \nu.
\end{equation}
TFH \cite{tfh85} fitted the cumulative distribution they obtained from a numerical resolution of
(\ref{eq:2.4}) for steady arrangements of aligned poles, by the expression $\pi^2\nu R =
\int_0^B\log (1.28 N\nu \pi^2/\left|B'\right|)\,dB'$ when $\left|B\right| \le B_{\mathrm{max}}$
\cite{tfh85}. Equations (\ref{eq:3.6d}) (\ref{eq:3.8}) show that 1.28 estimated from their numerical
pole distribution at $\left|B\right| \ll B_{\mathrm{max}}$ actually was a numerical approximation of
$4/\pi=1.273\ldots$ Figures \ref{fig:1} and \ref{fig:2} compare the analytical findings
(\ref{eq:3.7}) (\ref{eq:3.8}) to our own resolutions of (\ref{eq:2.4}), with $N=10,$ and 100,
respectively. The TFH fit is also displayed for illustration. The pole density
$P$ is defined for $\alpha \ge 1$ by
\begin{equation}
	\label{eq:3.8b}
	P((B_{\alpha}+B_{\alpha-1})/2) \equiv (B_{\alpha}-B_{\alpha-1})^{-1}, 
\end{equation}
in terms of the pole locations (with $B_0 = 0$ by convention); it is shown in Fig. \ref{fig:3} for $N=100,$ and compared with the continuous approximation (\ref{eq:3.6d}) and the TFH fit. Once the cumulative distribution is determined by
(\ref{eq:3.7}) (\ref{eq:3.8}) in the continuous limit, approximations $\tilde{B}_{\alpha}$ to the
discrete pole locations can be retrieved upon solving \cite{tfh85} 
\begin{equation}
\label{eq:3.9}
R(\tilde{B}_{\alpha}) = \alpha -1/2,\qquad \alpha=1,\ldots , N
\end{equation}
numerically ({\it e.g.,} by the Newton-Raphson method, with the ``exact" $B_{\alpha}$s as initial
guess!). The resulting crest shape
\begin{equation}
\label{eq:3.10}
\displaystyle
\tilde{\phi}(x) = -2\nu \sum_{\alpha = 1}^N \log\left(1+\frac{x^2}{\tilde{B}^2_{\alpha}}\right)
\end{equation}
is compared to the exact one (numerical) in Fig. \ref{fig:4} and to that issued from the continuous
approximation. The latter profile has
\begin{eqnarray}
 \phi_x & = & -\int_{-B_{\mathrm{max}}}^{B_{\mathrm{max}}} \frac{2\nu P(B)\,dB}{x-iB}\label{eq:23}\\
	& = & -\frac{1}{\pi}\,\sign(x)\,
\log\left(\frac{\sqrt{x^2/B_{\mathrm{max}}^2+1}+1}{\sqrt{x^2/B_{\mathrm{max}}^2+1}-1}\right)
\label{eq:24},
\end{eqnarray}
the second expression resulting from substitution of (\ref{eq:3.6d}) in (\ref{eq:23}), then a lucky
look at p. 591 of Ref.\cite{gr}.

As suggested by the form of (\ref{eq:23}), and confirmed by (\ref{eq:24}), $\phi_x(x)$ is most
simply deduced from $P(\pm ix)$ through contour integration in the complex $B$-plane. A further
integration by parts of (\ref{eq:24}) yields the continuous-approximation prediction for $\phi(x)$
(up to an additive constant):
\begin{equation}
	\label{eq:25}
	\phi(x) = -\frac{1}{\pi}\,\sign(x)\,B_{\mathrm{max}} \left(\sinh\xi\,
\log\frac{\cosh\xi+1}{\cosh\xi-1} + 2\xi\right),
\end{equation}
where $x = B_{\mathrm{max}}\,\sinh\xi$ (compare to (\ref{eq:3.3})). The integration constant was
selected in Fig. \ref{fig:4} to achieve good agreement
with the exact $\phi(x)$ for $\left|x\right|\rightarrow \infty$. Two final remarks: (i) $\nu$
disappeared as a factor in (\ref{eq:24}) as it should, because $\nu$ can be scaled
out; (ii) $\phi(x)$ is of the form $\nu N F(x/\nu N)$, and this scale-invariance shows that the
continuous approximation actually amounts to describing $\phi(x)$ at large distances compared to the
actual radius of curvature ($1/\int_{\tilde{B}_1}^{B_{\mathrm{max}}}4\nu P(B) dB/B^2 = o(\nu)$) of
the flame tip, when $N \gg 1$ ( that is, for large wrinkles).
\section{Mono-coalesced, periodic crest}
The following simple remark will allow us to solve (\ref{eq:2.6}), {\it i.e.,} in the case where all
the poles of $\phi_x$ are aligned along the imaginary $x$-axis ($\bmod~2\pi$). Because $P(B')$ 
still is an even function of $B'$, only the even parts (at fixed $B$) of $\coth ((B-B')/2)$ will
actually contribute to the integral over $B'$. Equation (\ref{eq:2.6}) may thus be re-written as 
\begin{equation}
\label{eq:4.1}
\displaystyle
\dashint_{-B_{\mathrm{max}}}^{B_{\mathrm{max}}} \frac{\nu P(B')
(1-\tanh^2(B'/2))}{\tanh(B/2)-\tanh(B'/2)}\,dB'=\sign(B),
\end{equation}
upon use of the known formula for the $\tanh(\cdot)$ of a difference, and neglect of a term
proportional to $\dashint P(B') \tanh(B'/2)\,dB' = 0.$ We now set
\begin{equation}
\label{eq:4.2}
\displaystyle
\tanh(\frac{B}{2}) = \tanh(\frac{B_{\mathrm{max}}}{2}) \sin \Phi,\quad-\frac{\pi}{2} \le \Phi \le
\frac{\pi}{2},
\end{equation}
converting (\ref{eq:4.1}) into
\begin{equation}
\label{eq:4.3}
\displaystyle
\dashint_{-\pi/2}^{\pi/2}\frac{2\nu P(B') \cos \Phi'}{\sin \Phi-\sin \Phi'}\,d\Phi' = \sign(\Phi),
\end{equation}
which is nothing but (\ref{eq:3.4}). Therefore the sought after pole-density is still given by
(\ref{eq:3.6c}), the only difference with the previous non periodic case being that $B$,
$B_{\mathrm{max}}$, and $\Phi$ are now related by (\ref{eq:4.2}) instead of (\ref{eq:3.3}). 

\begin{figure}
  \includegraphics[width=\textwidth]{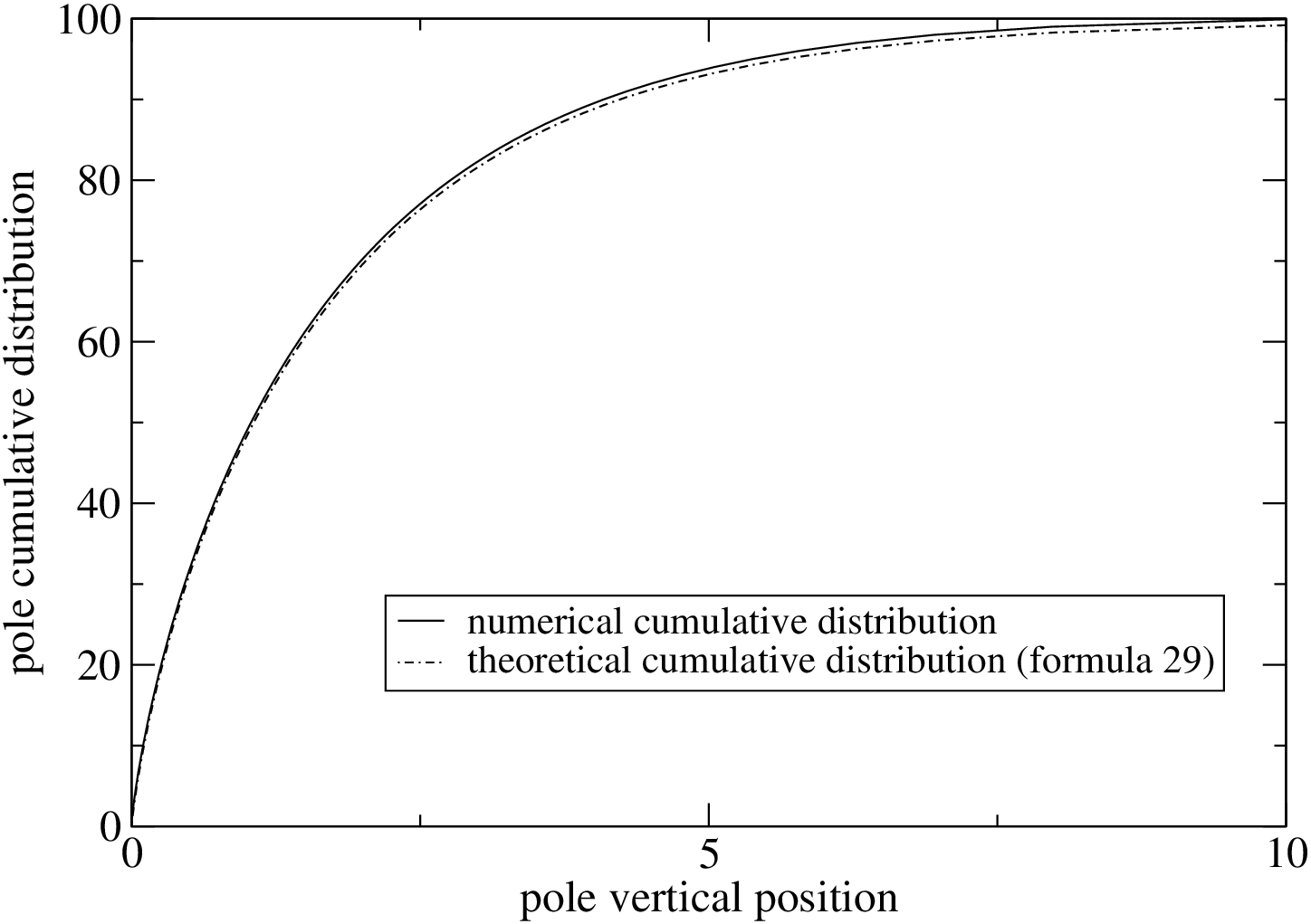}
  \caption{\label{fig:5}Numerical (solid line) vs analytical (eq. (\ref{eq:4.4}), dot-dashed line)
cumulative pole densities for a monocoalesced periodic crest, for $1/\nu = 199.5$, and $N = 100$
($=N_{\mathrm{opt}}(\nu)$). Only the upper hull of the numerical staircase is shown.}
\end{figure}

\begin{figure}
  \includegraphics[width=\textwidth]{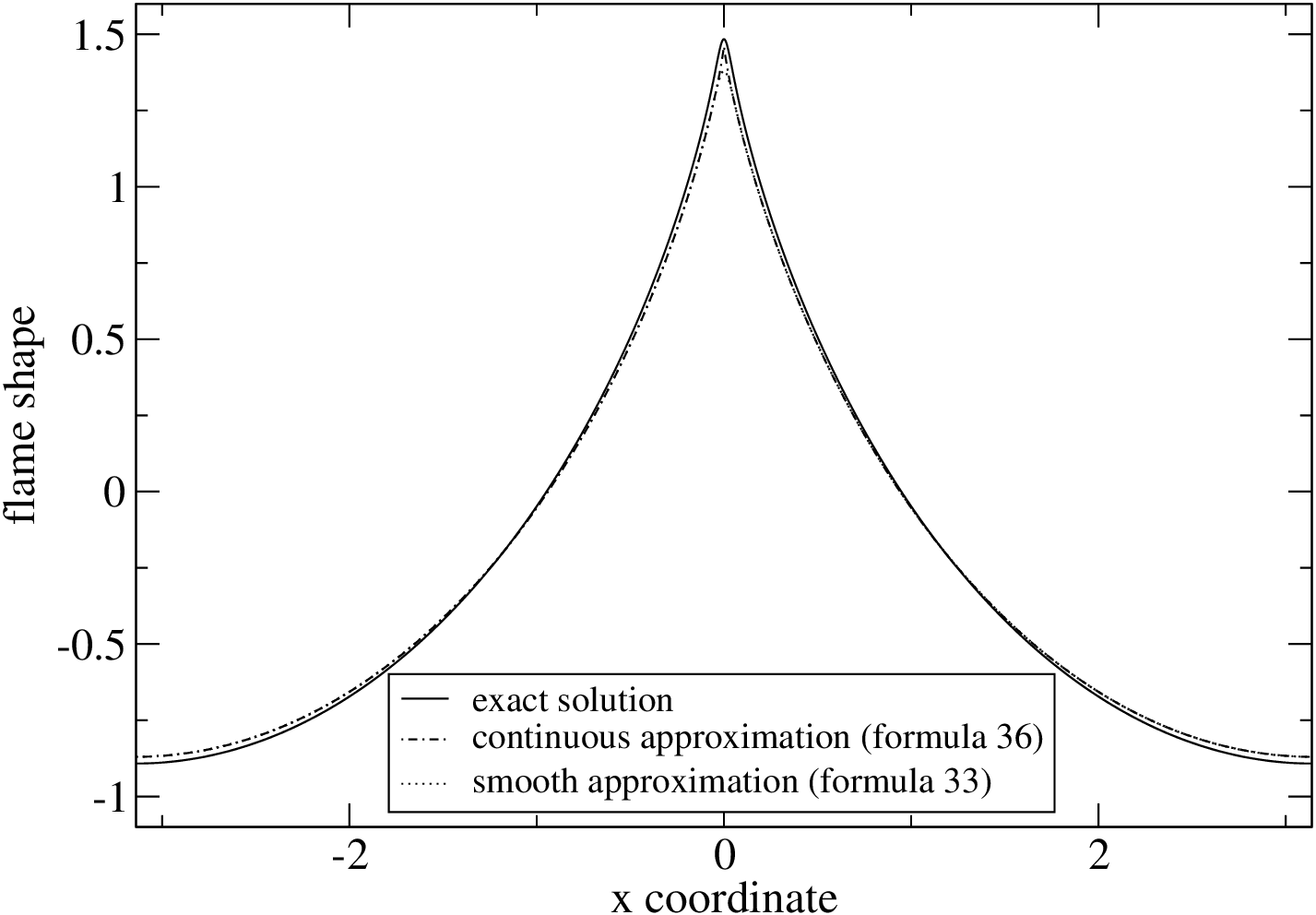}
  \caption{\label{fig:6}Shapes of a monocoalesced periodic flame with $1/\nu = 19.5$, $N = 10$
($=N_{\mathrm{opt}}(\nu)$): continuous approximation ((\ref{eq:4.11}), dot-dashed curve) vs exact
result (solid line) and smooth approximation ((\ref{eq:4.8}), dotted).}
\end{figure}

The new cumulative density $R(B)=\int_0^B P(B')\, dB'$ is given, after an integration by parts, by
\begin{eqnarray}
\pi^2\nu R(B) & = & \frac{1}{2}\log\frac{1+A \sin\Phi}{1-A \sin\Phi}\,\,\log\frac{1+ \cos\Phi}{1- \cos\Phi}\nonumber\\
	 &     &  + \int_0^{\Phi}\log\frac{1+A \sin\Phi'}{1-A \sin\Phi'}\,\frac{d\Phi'}{\sin\Phi'},
\label{eq:4.4}
\end{eqnarray}
$A \equiv \tanh(B_{\mathrm{max}}/2)$, whereby the normalization (\ref{eq:2.7}) requires
\begin{equation}
\label{eq:4.5}
\displaystyle
N\nu \pi^2 = \int_0^{\pi/2}\log\frac{1+A \sin\Phi}{1-A \sin\Phi}\,\frac{d\Phi}{\sin\Phi}.
\end{equation}
As the above integral turns out to be $\pi \arcsin A $ (p. 591 of \cite{gr}) the range of
$P(B)$, still given by
$R(B_{\mathrm{max}}) = N$, now satisfies
\begin{equation}
\label{eq:4.6}
\tanh(B_{\mathrm{max}}/2) = \sin(\pi N\nu)
\end{equation}
instead of (\ref{eq:3.8}). The latter and (\ref{eq:4.6}) coincide for $\pi N\nu \ll 1$, as do the
associated pole densities. The maximum $B_{\mathrm{max}}$ allowed by (\ref{eq:4.6}),
$B_{\mathrm{max}} = +\infty$, has $2N \nu = 1$ and $\cos \Phi\equiv 1/\cosh(B/2)$, whence $P(B)$
resumes the form
\begin{equation}
\label{eq:4.7}
\displaystyle
P(B) = \frac{1}{\pi^2\nu} \log\left(\coth\frac{\left|B\right|}{4}\right)
\end{equation}
obtained by TFH {\it via} Fourier transformations. The figure \ref{fig:5} compares our
predictions (\ref{eq:4.4}) and (\ref{eq:4.6}) with very accurate solutions of (\ref{eq:2.2}) for
$N=100$ and $2N\nu = 1$. Very good agreement is obtained even if $N$
is only moderately large, and carries over to the pole densities themselves. Again, approximate
solutions $\tilde{B}_{\alpha}$ can be retrieved from the analogue of (\ref{eq:3.9}), and an
approximate flame front shape from
\begin{equation}
\label{eq:4.8}
\displaystyle
\tilde{\phi}(x) = -2\nu \sum_{\alpha = 1}^N \log(1-\cos x \,\,\mathrm{sech}\, \tilde{B}_{\alpha}) + \mathrm{const}.
\end{equation}
Figure \ref{fig:6} shows of a comparison between (\ref{eq:4.8}), the exact flame shape obtained from
the exact (yet obtained numerically) $B_{\alpha}$s satisfying (\ref{eq:2.2}), and the curve deduced from the continuous approximation,
for which the flame slope $\phi_x(x)$ reads
\begin{equation}
\label{eq:4.9}
\displaystyle
\phi_x = -\nu \int \cot\left(\frac{x-i B}{2}\right) P(B)\,dB,
\end{equation}
again a real function because $P(-B)=P(B)$. With $P(B)$ given by (\ref{eq:3.6c}) (\ref{eq:4.2}) (\ref{eq:4.6}) the
above integral can be reduced to one available in p. 591 of \cite{gr} and yields (for
$-\pi\le x \le \pi$):
\begin{equation}
	\label{eq:4.10}
	\phi_x(x) = -\frac{1}{\pi}\,\sign(\xi)\,\log\frac{\cosh\xi+1}{\cosh\xi-1}, \qquad
\tan\frac{x}{2}\equiv A \sinh\xi
\end{equation}
thereby confirming that $\phi_x(x)$ is accessible from $P(B)$ by analytical continuation to $\pm i x$. In
particular, the TFH solution, eq. (\ref{eq:4.7}), has $\pi \phi_x =
-2\,\sign(x)\,\log\left|\cot x/4\right|$ and $\phi_{xx}(\pm\pi)=1/\pi$; more generally, $\phi_{xx}(\pm\pi)=A/\pi$ . A further
integration by parts yields
\begin{eqnarray}
	-i \pi\phi(x) =  \sign(x) \,\log\frac{1+ i A \sinh\xi}{1-i A
\sinh\xi}\,\log\frac{\cosh\xi+1}{\cosh\xi-1}\nonumber\\
+ 2\, \sign(x) \int_0^{\xi}\log\frac{1+ i A
\sinh\xi'}{1-i A \sinh\xi'}\,\frac{d\xi'}{\sinh\xi'}, \label{eq:4.11}
\end{eqnarray}
which cannot be evaluated in simple closed form, but may be compared to (\ref{eq:4.4}); of course
$\phi(x)$ is real when $x$ is, since the complex $\log(\cdot)$ in (\ref{eq:4.11}) also reads $2 i
\arctan(A\sinh\xi)=i x$. Note that
$\phi(x)$ has the form $F(x;N\nu)$, in the present units where the pattern is $2\pi$-periodic.
Adopting $\Lambda\neq 2\pi$ as wavelength would give $2\pi \phi =\Lambda
F(2\pi x/\Lambda;2\pi N\nu/\Lambda)$ with the same $F$. Accordingly, if $\nu N/\Lambda$ is
kept
fixed, $\phi_{xx}(\pm\Lambda/2)$ scales
like $1/\Lambda$ as it should for $\nu\rightarrow 0$, whereby halving the wavelength renders the
patterns less sensitive to noise (see
the {\it Introduction}).
\section{Flame speed from continuous pole-density}
Plugging (\ref{eq:4.9}) into (\ref{eq:2.3}) allows the wrinkling-induced increase in flame speed $V$
to be written as 
\begin{equation}
\label{eq:5.2}
\displaystyle
2 V = \nu^2 \int\!\!\!\!\int P(B)\,P(B') \langle\cot\frac{x-i B}{2}\cot\frac{x-i B'}{2}\rangle\,dB\,dB'.
\end{equation}
Although the one-variable integrals involved when squaring (\ref{eq:4.9}) are ordinary ones, they
may be written as principal parts. We next invoke the trigonometric identity $\cot a \cot b = -1 +
\cot(a-b) (\cot b - \cot a)$ and the average
\begin{equation}
\label{eq:5.3}
\displaystyle
\langle\cot\frac{x-i B}{2}\rangle = i~\sign(B)
\end{equation}
to transform (\ref{eq:5.2}) into
\begin{eqnarray}
2\frac{V}{\nu^2}  =   -\int\!\!\!\!\int P(B)\,P(B')\,dB dB'\rule{25mm}{0mm}\nonumber\\
                              + 2 \int\!\!\!\dashint\sign(B) P(B)\,P(B')
\coth\left(\frac{B-B'}{2}\right) dB' dB
\label{eq:5.4}
\end{eqnarray}
The first double integral ($=(\int P(B) dB)^2$) in (\ref{eq:5.4}) follows from the normalization 
(\ref{eq:2.7}), and is $(2N)^2$. The second one is obtained from (\ref{eq:2.6}) after multiplication
of both sides by $P(B) \sign(B) dB$ and subsequent integration over $B$: by (\ref{eq:2.7}), it is
$2N/\nu$. Thus the simple formula
\begin{equation}
\label{eq:5.5}
V = 2N \nu (1-N \nu)
\end{equation}
ensues; notice that it was obtained without having to solve (\ref{eq:2.6}). Actually (\ref{eq:5.4})
can be shown from (\ref{eq:2.2}) to hold whatever $N$ and $\nu$ \cite{joulin87}, again without
solving the pole-equations themselves.

In view of the accuracy of (\ref{eq:5.5}) one may inquire whether the solutions of (\ref{eq:2.5})
(\ref{eq:2.6}) satisfy the ``inviscid'' Sivashinsky equation, {\it i.e.} (\ref{eq:siveq}) with
$\nu=0$, in the steady cases. To show they do, for $x\neq 0$ at least, one may set $\mathcal{P}=P
\nu$
and $\mathcal{N}=N \nu$ to remove $\nu$ from (\ref{eq:2.5}) (\ref{eq:2.7}), then process the
Landau-Darrieus term of (\ref{eq:siveq}) as follows in the case of an isolated crest:
\begin{eqnarray}
 2i I(\phi) & = & \int \frac{4\,\mathcal{P}(B) \sign(B)}{x-iB}\,dB\nonumber\\
	    & = & \int \frac{2\,\mathcal{P}(B)}{x-iB}\,dB \,\,\dashint
\frac{2\,\mathcal{P}(B')}{B-B'}\,dB' + (B\leftrightarrow B')\nonumber\\
	    & = & \int\!\!\!\!\int \frac{4 i\, \mathcal{P}(B) \mathcal{P}(B')}{(x-iB) 
(x-iB')}\,dB\,dB' = i \phi_x^2,
\end{eqnarray}
where the notation ($B' \leftrightarrow B$) represents a second copy of the integral that precedes it, with
$B$ and $B'$ interchanged. The lines above successively use (\ref{eq:2.5}), acknowledge that
($B,B'$) are dummy variables of integration that may be interchanged, then employ (\ref{eq:23})
squared. Hence (\ref{eq:25}) satisfies (\ref{eq:siveq}) when $\nu=0$ and $\mathcal{N}$ is
prescribed, if $x \neq 0$. Thanks to (\ref{eq:5.4}), a similar analysis applies to (\ref{eq:2.6}),
provided $x \neq 0$ (mod $2\pi$).

Beside providing one with an exact $P(B)$, eq. (\ref{eq:3.6c}) shows that (\ref{eq:2.6}) admits a
continuum of solutions, for there exists nothing in (\ref{eq:2.7}) to tell one that $N$ ought to be
an integer; this will be commented later (see Sec. 8). One finally specializes (\ref{eq:2.6}) to
$B=B_{\mathrm{max}}$ to show that $N$ is constrained by $0 \le 2N\nu \le 1$, since
$\coth(B_{\mathrm{max}}-B)\ge 1$ (see also (\ref{eq:4.6})).

\section{Dynamics of supplementary pole-pairs}\label{sec:dynamics}

In 2000, Vaynblatt \& Matalon \cite{vm} addressed the {\it linear} stability of pole-decomposed
monocoalesced ``steady" solutions $-V t +\phi(x)$ to (\ref{eq:siveq}). Upon writing $\phi(x,t)+V t - \phi(x) \sim \exp(\omega t) \psi_{\omega}(x) \ll 1$ then analytically solving the linearised dynamics to get $\omega$ and $\psi_{\omega}(x)$, the authors of \cite{vm} identified two types of linear modes. The modes of type I describe how the $2N$ poles of $\phi_x(x)$ evolve when displaced by {\it infinitesimal} amounts from equilibrium; all those are stable ($\omega < 0$), but one that has $\omega = 0$ (see below). The modes of type II were interpreted \cite{vm,procaccia99} as resulting from $x$-periodic arrays of poles at $\pm i \infty$ that may spontaneously approach the real axis if $N$ is too small for the selected $\nu < 1$. The overall conclusion was thus: when  endowed with $2\pi$-periodic boundary
conditions, all the monocoalesced solutions are linearly unstable, except a single one that has $N=N_{\mathrm{opt}}(\nu)\equiv\lfloor(1+1/\nu)/2\rfloor \simeq 1/2\nu$
($\lfloor\cdot\rfloor \equiv$ integer part) and is neutrally stable ($\omega = 0$) against shifts along the
$x$-axis, the corresponding anti-symmetric eigenmode being $\psi_{0}(x) = \phi_x(x)$. For $N<N_{\mathrm{opt}}$, modes of type II can manifest themselves, two particularly dangerous ones corresponding to incipient secondary wrinkles centred on the main crests ($x=0$, $\bmod\, 2\pi$) or troughs ($x=\pi$, $\bmod\, 2\pi$).

When Neumann conditions  are employed instead, the aforementioned shifts are not allowed
any longer because $\psi_{0x} \neq 0$ at $x=0$ and $x=\pi$. Numerical integrations \cite{denet06} of (\ref{eq:siveq}) and
(\ref{eq:2.2})
evidence that there may then exist stable bi-coalesced patterns comprising an extra crest located at
$x=\pi$. Even though the steady $2\pi$-periodic patterns also satisfy (\ref{eq:siveq}) with Neumann conditions when properly shifted to have $\phi_x(0)=0=\phi_x(\pi)$ no stability analysis similar to \cite{vm} is yet available in this case; yet instabilities then necessarily require $N < N_{\mathrm{opt}}(\nu)$. Here we
address a restricted aspect of the problem, namely: we study how the previously determined monocoalesced ``steady"
solutions (\ref{eq:25}) (\ref{eq:4.11}) interact with extra pairs of poles. Since the free dynamics (\ref{eq:2.2}) conserves the total number of pole pairs at its $t=0$ value, it makes sense to consider $\phi(x,0)$s that involve them in a larger number ($N+n$) than the $N=O(1/\nu)$ ones retained in a steady profile $\phi(x)$. Each of the $n$ supplementary pairs at $x_m(t)\pm i y_m(t)$ contributes a perturbation $\phi_m(x,t) = \left<\phi_m\right> - 4\nu\sum_{j\ge 1} \exp(-j \left|y_m\right|) \cos(j(x-x_m))/j$ to the flame shapes (this follows from (\ref{eq:2.1}) via a term-by-term Fourier expansion \cite{tfh85}) and, as shown in \cite{procaccia99}, superposing $\phi_m$s can reproduce virtually any disturbance $\phi(x,0)-\phi(x)$. In the present formulation the only difference between Neumann and $2\pi$-periodic boundary conditions deals with the initial phases $x_m(0)$: whereas the former require the $x_m$s to be compatible with the $x \leftrightarrow -x$ and $\pi - x \leftrightarrow \pi + x$ symmetries, the latter do not.

Contrary to the more conventional normal-mode method (to which it is equivalent if $\left|y_m(0)\right| \gg 1$ \cite{procaccia99}), the pole approach can follow the disturbances when significant nonlinear effects set in\ldots if one is able to solve the $2N+2n$ coupled equations for the pole trajectories in the complex plane. The next remark somewhat simplifies the task. In the limits $N\gg 1$, $\nu \rightarrow 0^+$ and $\nu N = O(1)$ that led to (\ref{eq:2.6}), accounting for $n=O(1)$ extra pole pairs -- as is assumed here --  exerts only a small $O(\nu)$ perturbation on the $2N$ poles already aligned. Accordingly the distribution $P(B)$ of poles along the main alignments at $x=0$ ($\bmod\, 2\pi$) may be kept unchanged, and given by (\ref{eq:3.6c}) (\ref{eq:4.2}) (\ref{eq:4.6}),  when computing the motion of $2n$ supplementary ones. 

In the illustrative examples that follow only two extra poles ($n=1$) located at $\pm i\, y(t)\,(\bmod~2\pi)$, $y > 0$, then at $\pi\pm i y(t)$ are considered, to begin with.

\subsection{Extra-poles at $\bf x\simeq 0$ ($\bf\bmod~2\pi$)}\label{sec:dynamicsA}

When the two supplementary poles are located at $\pm i y(t)$ ($\bmod~2\pi$), their altitude $y(t)$ is determined from (\ref{eq:2.2}) -- within $O(\nu,1/N)$ fractional errors in the limits $N\gg 1$, $\nu \rightarrow 0^+$ and $\nu N = O(1)$ -- by the ODE
\begin{eqnarray}
\frac{dy}{dt} & = & \dashint \nu P(B') \coth\left(\frac{y-B'}{2}\right)\,dB' - 1,\label{eq:6.1}\\
                     & = & \frac{2}{\pi} \arcsin(\sin(\pi N\nu) \coth(y/2))-1,\quad \left|y\right|\ge B_{\mathrm{max}}, \label{eq:6.1b} 
\end{eqnarray}
where $P(B)$ is the {\it same} as given by (\ref{eq:3.6c}) (\ref{eq:4.2}) and (\ref{eq:4.6}), to
leading order, and leads to the closed form (\ref{eq:6.1b}) on integration \cite{gr}; for $\left|y\right| \le  B_{\mathrm{max}}$, $dy/dt = 0$ by (\ref{eq:2.6}). Therefore, whenever $0 < 1-2\nu N = O(1)$ and $\nu \rightarrow 0^+$, any initial $y(0) > B_{\mathrm{max}}$
will ultimately lead to $y(+\infty) = B^+_{\mathrm{max}}$, thereby adding one new incomer to the
already present continuum. Put in words: if $2\nu N< 1$ initially, the main pattern is unstable
to disturbances with poles at $\pm i\,y(t)$, and the latter process tends to make $2N \nu$
approach 1 from below.

Periodic boundary conditions would allow the supplementary pair to be initially off the $x$-axis, say at $x(0)\pm i y(0)$ with $0 < x(0) < \pi$ ($\bmod~2\pi$). The ``horizontal'' attraction by the main pole condensation at $x=0$ ($O(\nu)$, actually) \cite{tfh85} will make $x(t)$ decrease, while $y(t)$ still does if $2N \nu < 1$. Ultimately, the extra pole pair will join the main pole alignment (in finite time), and the previous conclusion is qualitatively unchanged: the process makes $N$ increase by one. When Neumann conditions are adopted, however, at least {\it two} pairs $\pm x(t)\pm i y(t)$ are needed if $x(t) \neq 0$,  to meet the requirement of symmetry about $x=0$, and two possibilities are encountered as to their fate. In the first instance, corresponding to not-too-small $x(0)$s and moderate values of $y(0)$, the process is qualitatively the same as above, except that 2 pairs simultaneously join the main condensation at $\left|y\right| < B_{\mathrm{max}}$, thereby making $N$ increase by 2. If $x(0)$ is small and $y(0)$ well above $B_{\mathrm{max}}$, the horizontal mutual attraction between the pair members may make them hit the $x=0$ axis at such a finite time $t_c$ that $y(t_c) > B_{\mathrm{max}}$; this is best shown from (\ref{eq:2.2}) specialized to $x(t) \ll 1$, whereby $dx/dt \simeq -\nu/x$ then $x^2(t)+2\nu(t-t_c) \simeq 0$ for $t \lesssim t_c$. The double pole thus formed at $i y(t_c)$ then instantly splits into two simple ones lying {\it on} the $x=0$ axis at $y(t)-y(t_c) \sim \pm (t-t_c)^{1/2}$, leading to a subsequent dynamics that ultimately ends like at the beginning of this subsection if $2 N \nu < 1$.

\subsection{Extra-poles at $\bf x\simeq \pi$ ($\bf\bmod~2\pi$)}

In case the supplementary poles are located at $\pi\pm i y(t)$  eq. (\ref{eq:6.1}) is  replaced by
\begin{equation}
\label{eq:6.2}
\displaystyle
\frac{dy}{dt}=\frac{2}{\pi} \arcsin(\sin(\pi N \nu) \tanh(y/2)) - 1+\nu \coth y,
\end{equation}
since $\coth(u+i\,\pi/2)=\tanh u$. Even though $\nu \ll 1$ the last term in (\ref{eq:6.2}),
stemming from the interaction of the extra-pole with its complex conjugate, cannot be simply discarded, for
otherwise (\ref{eq:6.2}) would not be uniformly valid if $y$ gets small. According to (\ref{eq:6.2}), 
any initial $y(0)$ indeed ultimately leads to $y(+\infty)
=\nu + o(\nu)$ and to a small $(O(\nu))$ stable disturbance centred at  $x=\pi$ ($\bmod~2\pi$) whenever $2N\nu < 1$.
Although the main pattern's curvature $\phi_{xx}(\pi)> 0$ is $O(1)$, and the $O(\nu)$ contribution
to $\phi(x,t)$ of the extra pole-pair is small, it is nevertheless enough \cite{denet06} to render
the flame shape $\phi(x,t)$ concave downward at $x=\pi$; as shown in \cite{denet06}, this occurs
as soon as the extra poles enter a thin strip about the real axis, $\left|y\right|\lesssim
\sqrt{4\pi\nu}$. Incorporating $O(N)$ extra pairs will also
do, but the process of dynamical trough splitting is not within reach of such ODEs as
(\ref{eq:6.2}) when $n=O(N)$. The structure of 2-crested steady patterns with $n=O(N)$ will be studied in Sec.\ref{sec:bico}.

Like in \ref{sec:dynamicsA} one might begin generalizing the present discussion by envisaging a single pair of extra poles off the $x=\pi$ axis, but this is already covered in the preceding paragraphs: if $0 < x(0) < \pi$ the pair ultimately joins the poles at $x=0$ ($\bmod~2\pi$). It is more revealing to consider two such pairs at $\pi \pm x(t) \pm i y(t)$  with $x(t)$ ``small enough'', in a way compatible with Neumann conditions, because something new appears. Comparatively large $x(0)$s will clearly lead to pairs that ultimately stick at $x=0$ ($\bmod~2\pi$), because their mutual horizontal attraction could not oppose that of the main alignments. The other extreme of very small $x(0)$s again leads to the formation of double poles at some $\pi \pm i y(t_c)$, then a subsequent evolution of the two pairs $\pi \pm i y_{1,2}$ {\it along} the line $x=\pi$ ($\bmod~2\pi$) until they settle at $O(\nu)$ distances to the real axis if $2 N \nu <1$. The important conclusion is that stable 2-crest patterns exist when Neumann conditions are used and $2 N \nu < 1$.

By continuity there exist separating trajectories $S_{ \pm}$, such that none of the above behaviours is observed if the pole pairs initially sit on them. The lines $S_{\pm}$ lead the two supplementary pairs towards an unstable equilibrium, a result of a competition between attraction by the main pole population at $x=0$ ($\bmod~2\pi$), and the mutual attractions/repulsions among the pair members. For $\nu \ll 1$, and $N \nu = O(1)$, using the steady version of (\ref{eq:2.2}) and the pole-density given by (\ref{eq:3.6c}) (\ref{eq:4.2}) (\ref{eq:4.6}), one can show that such equilibriums correspond to $x(+\infty) = \pm (2\pi\nu/A)^{1/2} +\cdots$ and $y(+\infty) = \pm \nu+\cdots$ to leading order, again with $A=\tanh(B_{\mathrm{max}}/2) = \sin(\pi N \nu)$. This shows that there exist even more general steady solutions than considered elsewhere in the paper and in the literature (except in \cite{denet06} where a similar conjecture was made on a numerical basis). One could have included other pairs as well, some of which along the $x=\pi$ ($\bmod~2\pi$) axis.

Our last remark is to again stress that the free dynamics (\ref{eq:2.2}) conserves the total number of poles (if finite). By the same token, allowing this number to vary with time is a means to study a {\it forced} version of the Sivashinsky equation: {\it adding} a pair of poles $x_m\pm i y_m$ at $t=t_m$ amounts to accounting for a term $\phi_m(x)\, \delta(t-t_m)$ in the right-hand side of (\ref{eq:siveq}), and combining many $\phi_m$s with various phases (as to vary their signs), amplitudes ($\simeq -4\nu\exp(-\left|y_m\right|)$ if $\left|y_m\right| \gg 1$) and times of implantation ($t_m$) could help one investigate the response of flames to a rich class of weak random noises. We understand that a similar proposal was developed about the ``kicked'' Burgers equation \cite{bfk}, {\it i.e.}, (\ref{eq:siveq}) without the integral term in the one-dimensional case.

\section{Bi-coalesced periodic patterns}\label{sec:bico}
% 
% For reasons that shall become clearer after Eq. (\ref{eq:7.8}), we provisionally rename the variables ($\Phi, \Phi'$) in (\ref{eq:3.4}) as ($\theta, \theta'$) then change the variable from $\theta$ to $\sigma$, with
% %
% \begin{equation}
% \label{eq:7.1}
% \displaystyle
% \sin \theta = \frac{(1+\varepsilon) \sin\sigma}{1+\varepsilon \sin^2\sigma}
% \end{equation}
% %
% so that $-\pi/2 \le \sigma \le \pi/2$ and $\sign(\theta) = \sign(\sigma)$. In (\ref{eq:7.1}), $0 \le
% \varepsilon \le 1$ is an as yet unspecified real (and arbitrary) parameter. Given that $P$ is an
% even function of its argument, (\ref{eq:7.1}) transforms (\ref{eq:3.4}) into
% %
% \begin{eqnarray}
% \dashint_{-\pi/2}^{\pi/2}\frac{d\sigma' \cos\sigma'2\nu P(\sigma')}{\sin\sigma-\sin\sigma'}\rule{33mm}{0mm}\nonumber\\
% + \varepsilon\sin\sigma\int_{-\pi/2}^{\pi/2}\frac{d\sigma' \cos\sigma'2\nu
% P(\sigma')}{1-\varepsilon^2\sin^2\sigma \sin^2\sigma'} = \sign(\sigma),
% \label{eq:7.2}
% \end{eqnarray}
% %
% that is admittedly more complicated than (\ref{eq:3.4}), but only superficially so: indeed, since
% the solution to (\ref{eq:3.4}) is known, $2\nu\pi^2P = \log(\cot^2(\theta/2))$ in the present
% notation, that of (\ref{eq:7.2}) is immediately available {\it via} (\ref{eq:7.1}).

We now take up the structure of ``steady" $2\pi$-periodic solutions of (\ref{eq:siveq}) that
would have $N$ pairs of poles $i\,B_{\alpha}$ $(\bmod~2\pi)$, $\alpha = \pm 1, \pm 2, \ldots, \pm N$,
and $n = O(N)$ other pairs at $\pi + i~ b_{\gamma}$ $(\bmod~2\pi)$, $\gamma = \pm 1, \pm 2, \ldots, \pm n$.
For brevity, we will say that the pole-alignments reside ``at" $x=0$ or $x=\pi$,
respectively, like the two crests per-cell they correspond to. Because
$\coth(u+i\,\pi/2)\equiv\tanh(u)$, the steady versions of (\ref{eq:2.2}) corresponding to such
bi-coalesced flame patterns read as
\begin{eqnarray}
\nu \mathop{\sum_{\beta=-N}^{N}}_{\beta \neq \alpha}\coth\left(\frac{B_{\alpha}-B_{\beta}}{2}\right)\rule{30mm}{0mm}\nonumber\\ 
+ \nu \sum_{\delta=-n}^{n}\tanh\left(\frac{B_{\alpha}-b_{\delta}}{2}\right) = 
\sign(B_{\alpha}),\label{eq:7.3a}\\
\nonumber\\\nonumber\\
\nu\mathop{\sum_{\delta=-n}^{n}}_{\delta\neq\gamma}\coth\left(\frac{b_{\gamma}-b_{\delta}}{2}
\right)\rule{34mm}{0mm}\nonumber\\ 
+ \nu \sum_{\beta=-N}^{N}\tanh\left(\frac{b_{\gamma}-B_{\beta}}{2}\right) = 
\sign(b_{\gamma}).\label{eq:7.3b}\\\nonumber
\end{eqnarray}
In the distinguished limits $\nu\rightarrow 0^+$, $N\nu = O(1)$, $n\nu=O(1)$, the poles at $x=0$ and
$x=\pi$ get densely packed (at the scale of the wavelength), with densities $P(B)$ and $p(b)$,
respectively. Both $P$ and $p$ will be nonnegative and, in general, compactly supported:
$P(\left|B\right| \ge B_{\mathrm{max}}) = 0 = p(\left|b\right| \ge b_{\mathrm{max}})$. The ranges
$B_{\mathrm{max}}$ and $b_{\mathrm{max}}$ are to be found as part of the solutions to the
continuous versions of (\ref{eq:7.3a}) and (\ref{eq:7.3b}):
\begin{eqnarray}
\dashint\nu P(B')\coth\left(\frac{B-B'}{2}\right)\,dB' \rule{30mm}{0mm}\nonumber\\
+ \int\nu p(b')\tanh\left(\frac{B-b'}{2}\right)\,db' = \sign(B),\label{eq:7.4a}\\
\nonumber\\\nonumber\\
\dashint\nu p(b')\coth\left(\frac{b-b'}{2}\right)\,db' \rule{36mm}{0mm}\nonumber\\
+ \int\nu P(B')\tanh\left(\frac{b-B'}{2}\right)\,dB' = \sign(b),\label{eq:7.4b}
\end{eqnarray}
that are valid for $\left|B\right|\le B_{\mathrm{max}}$ and $\left|b\right| \le b_{\mathrm{max}}$,
respectively. To restore some symmetry we set
\begin{eqnarray}
\tanh(B/2) & = & A \sin\Phi, \,\, A\equiv \tanh(B_{\mathrm{max}}/2) \le 1,\label{eq:7.5a}\\
\tanh(b/2) & = & a \sin\varphi, \,\,\,\, a\equiv \tanh(b_{\mathrm{max}}/2) \le 1,\label{eq:7.5b}
\end{eqnarray}
in (\ref{eq:7.4a}) (\ref{eq:7.4b}), then acknowledge that both $P(\cdot)$ and $p(\cdot)$ are even
functions, which allows one to suppress some odd parts of the integrands, viewed as functions of
$\Phi'$ (or $\varphi'$) at fixed $\Phi$ (or $\varphi$). Some cumbersome algebra ultimately
transforms (\ref{eq:7.4a}) and (\ref{eq:7.4b}) into
\begin{eqnarray}
\dashint\frac{2\nu P(\Phi')\cos\Phi'}{\sin\Phi - \sin\Phi'}\,d\Phi' + A a \sin\Phi\nonumber\rule{20mm}{0mm}\\
\times \int\frac{2\nu p(\varphi')\cos\varphi'}{1-A^2a^2 \sin^2\varphi' \sin^2\Phi}\,d\varphi' = \sign(\Phi)\label{eq:7.6a}\\\nonumber\\\nonumber\\
\dashint\frac{2\nu p(\varphi')\cos\varphi'}{\sin\varphi - \sin\varphi'}\,d\varphi' + A a \sin\varphi\nonumber\rule{20mm}{0mm}\\
\times \int\frac{2\nu P(\Phi')\cos\Phi'}{1-A^2a^2 \sin^2\Phi' \sin^2\varphi}\,d\Phi' = \sign(\varphi)\label{eq:7.6b}
\end{eqnarray}
where all the variables ($\Phi, \Phi'$), ($\varphi, \varphi'$) are now taken in the common
$[-\pi/2, \pi/2]$ range. One may thus adopt a common notation for them, $(\sigma, \sigma')$ say, in both
(\ref{eq:7.6a}) and (\ref{eq:7.6b}) and subtract the results to eliminate the $\sign(\cdot)$
functions in the right-hand sides. This produces an homogeneous equation for the difference
$P(\cdot)-p(\cdot)$, of which one obvious solution is $P-p \equiv 0$. Hence the important result: if
$P=p$ is indeed a viable solution, then 
\begin{eqnarray}
P(B) & = & J\left(\sigma=\arcsin\left(\frac{\tanh B/2}{A}\right)\right),\label{eq:7.7a}\\
p(b) & = & J\left(\sigma=\arcsin\left(\frac{\tanh b/2}{a}\right)\right),\label{eq:7.7b}
\end{eqnarray}
where $J(\sigma)$ is the {\it same} function for both. The even $J(\cdot)$ function itself is then found
from (\ref{eq:7.6a}) or (\ref{eq:7.6b}) to satisfy
\begin{eqnarray}
\dashint_{-\pi/2}^{\pi/2}\frac{2\nu J(\sigma')\cos\sigma'}{\sin\sigma - \sin\sigma'}\,d\sigma' + A a
\nonumber\rule{25mm}{0mm}\\
\times \int_{-\pi/2}^{\pi/2}\frac{2\nu J(\sigma')\sin\sigma\cos\sigma'}{1-A^2a^2 \sin^2\sigma'
\sin^2\sigma}\,d\sigma' = \sign(\sigma).\label{eq:7.8}
\end{eqnarray}
Further changing the independent variable to $\theta$, with
\begin{equation}
 \sin\theta  =  \frac{(1+A a) \sin\sigma}{1+A a \sin^2\sigma},\label{eq:7.10}
\end{equation}
fortunately converts the seemingly hopeless (\ref{eq:7.8}) into a form equivalent to the already solved Eq. (\ref{eq:4.3}), $\theta$
playing the part that the former $\Phi$ did there (most easily shown by starting from (\ref{eq:4.3})).
Accordingly, the solution to (\ref{eq:7.8}) is available in terms of the already found pole-density
pertaining to the isolated crests, then the monocoalesced ones: from (\ref{eq:3.6c}) one indeed has
\begin{equation}
2\nu J(\sigma) = \frac{1}{\pi^2} \log\frac{1+\cos\theta}{1-\cos\theta}, \label{eq:7.9}
\end{equation}
with $\theta$ defined in (\ref{eq:7.10}). As said earlier, eqs. (\ref{eq:7.7a}) (\ref{eq:7.7b}), $P(B)$ is immediately retrieved upon setting $\sin\sigma = \tanh(B/2) \coth(B_{\mathrm{max}}/2)$ in (\ref{eq:7.10}) (\ref{eq:7.9}); same operation to get $p(b)$
from $J(\sigma)$, upon setting $\sin\sigma = \tanh(b/2) \coth(b_{\mathrm{max}}/2)$ in (\ref{eq:7.9})
(\ref{eq:7.10}).

\begin{figure}
  \includegraphics[width=\textwidth]{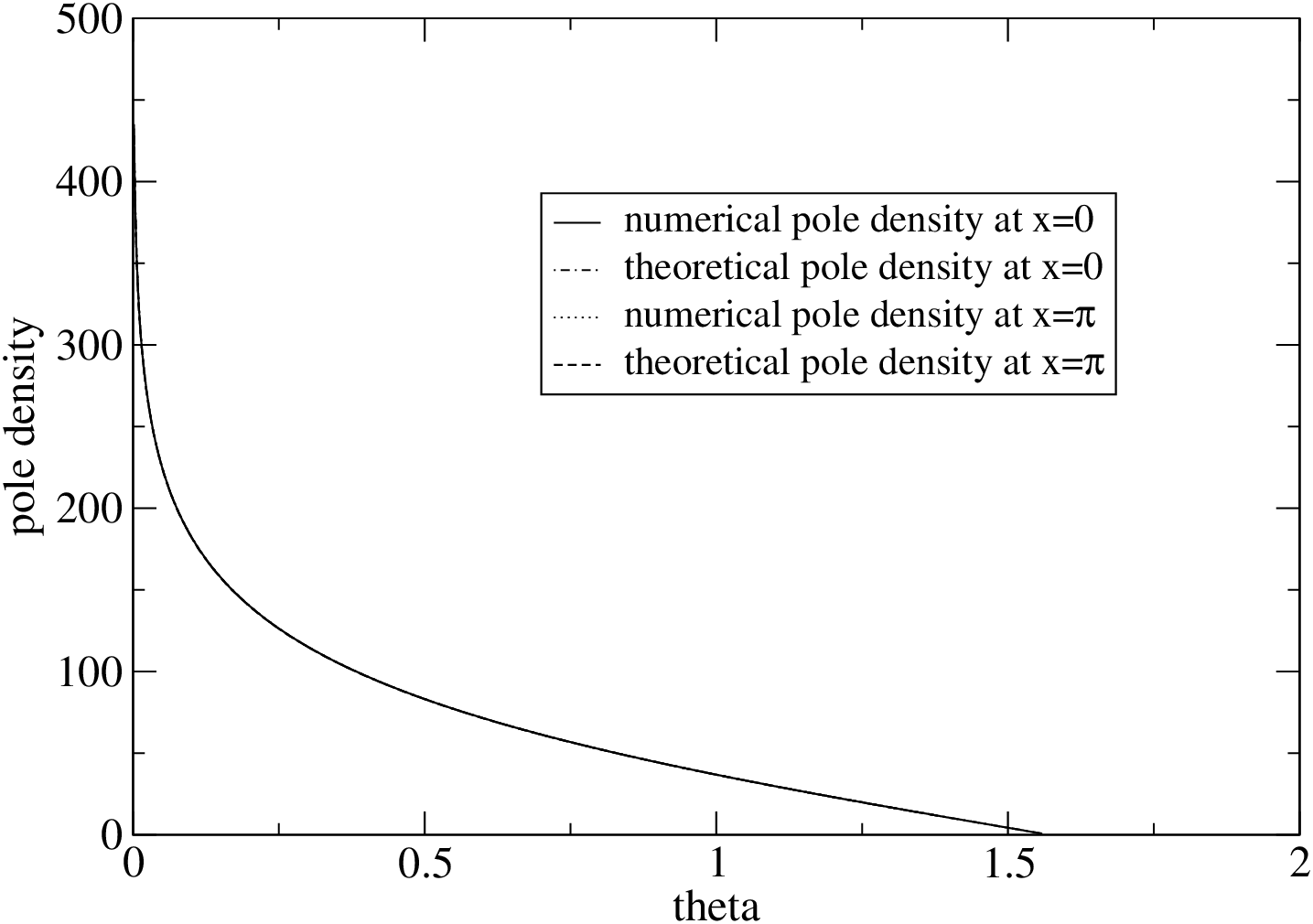}
  \caption{\label{fig:8}Theoretical pole densities $P(B)$ (resp. $p(b)$) plotted as dot-dashed 
or dashed lines vs $\theta$, eq. (\ref{eq:7.9}), with $\sin\sigma$ replaced by $(\tanh B/2)/A$
(resp. $(\tanh b/2)/a$). The solid and the dotted lines are the numerical pole densities. All are
for a bicoalesced periodic flame with $N=200$, $n=100$, $1/\nu=600.5.$}
\end{figure}

The first step to get $B_{\mathrm{max}}$ and $b_{\mathrm{max}}$ again is to compute the cumulative
pole-densities. For example $R(B) = \int_0^B
P(B')\,dB'$ is computed as follows from (\ref{eq:7.9}) (\ref{eq:7.10}):
\begin{eqnarray}
2\pi^2\nu R(B)\!\!\! & = &\!\!\!\! \int_0^{\Phi}\log\left(\frac{1+\cos\theta'}{1-\cos\theta'}\right) \frac{2A \cos\Phi' d\Phi'}{1-A^2\sin^2\Phi'},\label{eq:7.11}\\
                              & = & \!\!\log\frac{1+A\sin\Phi}{1-A\sin\Phi} \log\frac{1+\cos\theta}{1-\cos\theta}\nonumber\\
	          &    & + \,2\int_{0}^{\theta(\Phi)}\frac{d\theta'}{\sin\theta'}\log\frac{1+A\sin\Phi'}{1-A\sin\Phi'},\label{eq:7.12}
\end{eqnarray}
again with the understanding that $\Phi$ (or $\Phi'$) is viewed as a function of $\theta$ (or
$\theta'$) via (\ref{eq:7.10}), and conversely; (\ref{eq:7.11}) is obtained from the definition of
$R(B)$ upon setting $\tanh(B/2)=A\sin\Phi$, and (\ref{eq:7.12}) results from an integration by
parts. The cumulative density pertaining to $p(b)$ is obtained in the same way from (\ref{eq:7.9})
(\ref{eq:7.10}), now thanks to $\tanh(b/2)=a \sin\varphi$: the result
is like (\ref{eq:7.12}), except for the substitutions $A \rightarrow a$, $\Phi \rightarrow \varphi$,
$B \rightarrow b$, $R(B) \rightarrow r(b)$. The normalisations $R(B_{\mathrm{max}}) = N$ and
$r(b_{\mathrm{max}}) = n$ thus impose the two conditions
\begin{eqnarray}
N\nu\pi^2 & = & \int_0^{\pi/2}\frac{d\theta}{\sin\theta}\log\frac{1+A \sin\Phi}{1-A \sin\Phi},\label{eq:7.13a}\\
n\nu\pi^2 & = & \int_0^{\pi/2}\frac{d\theta}{\sin\theta}\log\frac{1+a \sin\varphi}{1-a \sin\varphi},\label{eq:7.13b}\\\nonumber
\end{eqnarray}
that may be compared to the former equation (\ref{eq:4.5}), and reduce to it when $A a = 0$.
Although we could not compute the above normalization integrals in closed forms, this can
be done numerically without difficulty to get $A$ and $a$ as function of $N\nu$ and $n\nu$ (or
conversely). Note that $N \ge n$ is equivalent to $A \ge a$. $N>n$ also implies that $R(\cdot)>
r(\cdot)$ when both are evaluated at the same argument, Fig. \ref{fig:8}.

Before closing this section, it remains to compare the above predictions to direct numerical
resolutions of (\ref{eq:7.3a}) (\ref{eq:7.3b}), by the Newton-Raphson method. This is done in Figs
\ref{fig:7} and \ref{fig:8}. Figure \ref{fig:7} will hopefully convince the
reader that both $P(B)$ and $p(b)$ can be expressed in terms of the single function $J(\sigma)$
given by (\ref{eq:7.9}).

Now that the pole densities are available, one may try to compute the corresponding increase in
flame speed, $V$, from (\ref{eq:7.4a}) (\ref{eq:7.4b}) without solving them (like in Sec. 5), to
produce
\begin{equation}
\label{eq:7.14}
V = 2\nu(N+n) (1-(N+n)\nu);
\end{equation}
this simple formula reduce to (\ref{eq:5.5}) if $n=0$, and could have been deduced directly from the
discrete pole-equations, without solving them. The sum $N+n$ plays the part $N$ did for
monocoalesced patterns and, as is shown upon specializing (\ref{eq:7.4a}) to $B=B_{\mathrm{max}}$,
has to satisfy $2(N+n) \nu \le 1$.

As mentioned earlier, the flame slope $\phi_x(x)$ pertaining to the continuous approximation(s) can
be obtained directly from the corresponding pole-density(ies) via an analytical continuation from
the real $B$ (or $b$) axis to $\pm i\,x$. Using the same procedure here gives, for $0\le x \le \pi$:
\begin{eqnarray}
\phi_x & = &-\frac{1}{\pi}\,\sign(\bar{x}-x)\,\log\frac{\cosh\psi+1}{\cosh\psi-1}\label{eq:7.15a}\\
\sinh\psi & = & \frac{(1+A a) \tan x/2}{A-a \tan^2x/2}\label{eq:7.15b}
\end{eqnarray}
for bicoalesced flames, $\bar{x}$ being the point where $\sinh^2\psi \rightarrow \infty$ in
(\ref{eq:7.15b}) and, therefore, $\phi_x(\bar{x}) = 0$:
\begin{equation}
\label{eq:7.16}
\bar{x} = 2 \arctan\sqrt{\frac{\tanh B_{\mathrm{max}}/2}{\tanh b_{\mathrm{max}}/2}}.
\end{equation}
At the flame front trough, $\kappa=\phi_{xx}(\bar{x})=2(A+a)/\pi(1+Aa) > A/\pi$: the corresponding critical noise amplitude $\mu_c(\kappa)$ needed to trigger the appearance of subwrinkles markedly exceeds that pertaining to monocoalesced fronts (see Sec. 1). Two extra pole-pairs initially placed at
the points $\pm\bar{x}\pm i\nu$ (mod $2\pi$) would stay there in unstable equilibrium. There exist separating trajectories
$S_{\pm}$ passing through them, which delineates the basins of attraction of the main pole
condensations at $x=0$ or $x=\pi$. Only the trajectories of initially remote extra-poles that are
close enough to $S_{\pm}$ will enter the $O(\sqrt{\nu})$ strip adjacent to the $B=0$ axis  where their
direct influence on the main pattern becomes visible \cite{denet06}. As seen from the real axis, the
process then manifests itself as extra sharp sub-wrinkles seemingly ``emitted'' suddenly at $x\simeq\pm\bar{x}$ (mod $2\pi$)
before travelling to one of the main cusps where they eventually join a main condensation. The $N/n$-dependent
shape of such separating trajectories thus controls the fate of ``supplementary'' poles of whatever origin, initial conditions or forcing; this will be exploited elsewhere, though one can already confirm that stable 2-crest steady patterns with $n=O(N) \sim 1/\nu$ exist if Neumann conditions are employed. With $2\pi$-periodic conditions these are unstable even if $2\nu (N+n) =1$, as is seen by considering initial conditions where the $2n$ poles are slightly shifted to the left of $x=\pi$ ($\bmod\, 2\pi$): both crests will ultimately merge.

\begin{figure}
  \includegraphics[width=\textwidth]{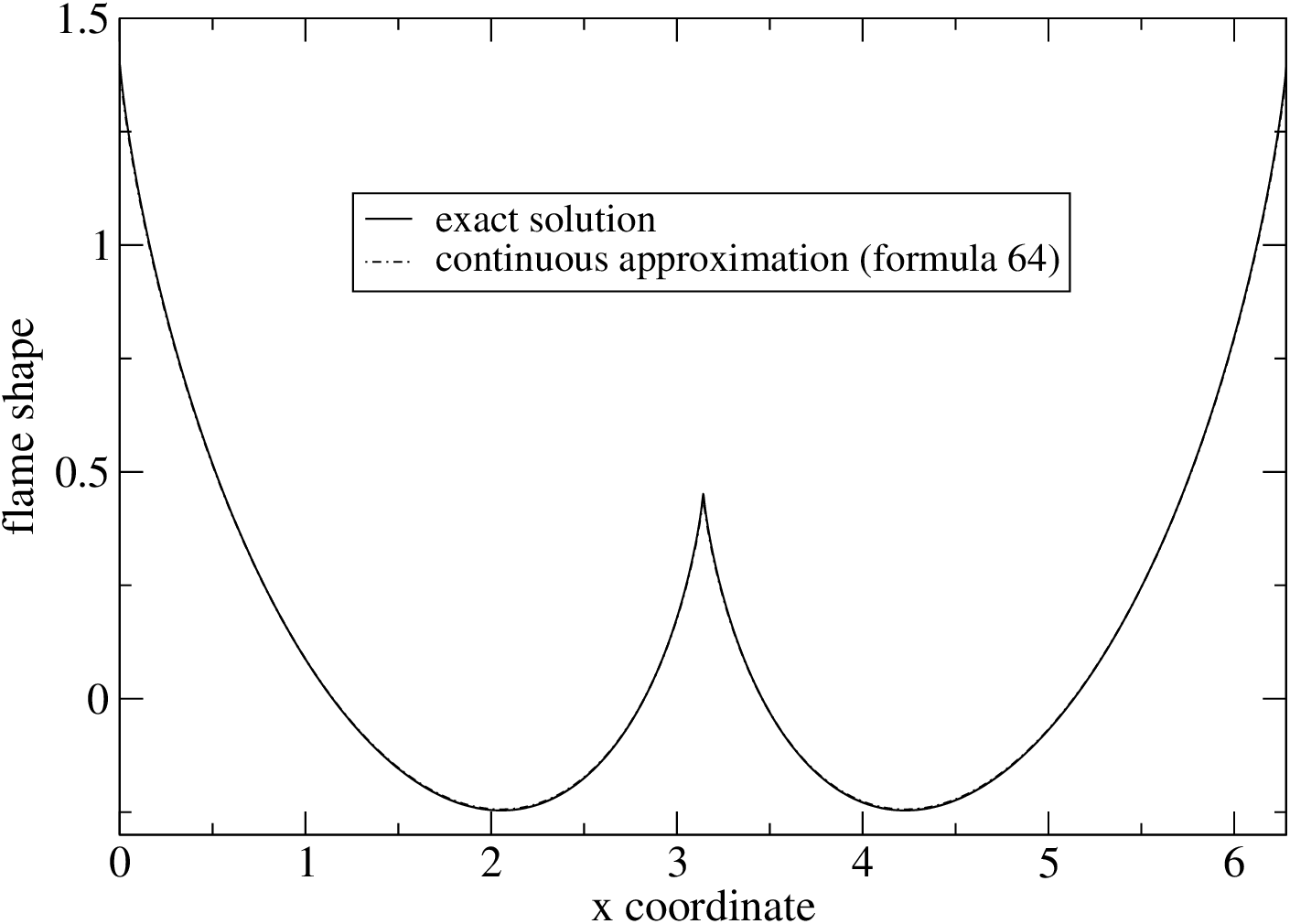}
  \caption{\label{fig:9}Shapes of a bicoalesced flame with $1/\nu = 199.5$, $N = 80$, $n = 20$:
exact (solid line) vs continuous approximation (from integration of (\ref{eq:7.15a}),
dot-dashed).}
\end{figure}

Comparisons with accurate numerical resolutions of the pole equations (\ref{eq:7.3a}) (\ref{eq:7.3b}) are again good, Fig.\ref{fig:9}. For $\nu = 1/199.5$, $N=80$, $n=20$ they yield $\bar{x}=2.053973$ whereas our prediction (\ref{eq:7.16}), with $A$ and
$a$ iteratively obtained from the normalization conditions (\ref{eq:7.13a}) (\ref{eq:7.13b}), gives $\bar{x}=2.053888$. Like
$R(B)$ and $r(b)$, $\phi(x)$ cannot be obtained in closed form, yet is readily accessible
numerically. Also, if $A=a$, elementary trigonometry shows that the predicted flame slope (\ref{eq:7.15a}) resumes
the result (\ref{eq:4.10}), up to a two-fold reduction in $x$- and $B_{\mathrm{max}}$-scales.
\section{Concluding remarks \& open problems}
The above analyses may convey the feeling that the pole densities obtained so far have a family
likeness, which is true because they were all deduced from the solution (\ref{eq:3.6c}) pertaining to isolated crests via adequate changes of
independent variable. Whereas (\ref{eq:3.6c}) itself basically follows from standard Fourier
analysis combined with a lucky re-summation of the series thus obtained (\ref{eq:3.6a}), it would be interesting to
understand why the changes of variable (\ref{eq:4.2}) then (\ref{eq:7.10}) work so well. Admittedly the integral equation (\ref{eq:4.1}) bears some
formal resemblance with (\ref{eq:2.5}), which guided us to propose the new variable (\ref{eq:4.2}); but
that introduced in (\ref{eq:7.10}) looks more strange to us, and was actually discovered by trial-and-error after the 'resolvent' integral equation (\ref{eq:7.8}) is obtained. Yet (\ref{eq:7.10}) unlikely solely comes ``out of the blue''. In effect,
one may note that (\ref{eq:7.10}) is equivalent to $\tanh(\beta) =
\tanh(\beta_{\mathrm{max}}) \sin\theta$ if one defines $\tanh^2(\beta_{\mathrm{max}}) \equiv
\tanh(B_{\mathrm{max}}/2) \tanh(b_{\mathrm{max}}/2)$ and sets $\tanh(\beta/2) =
\tanh(\beta_{\mathrm{max}}/2) \sin\sigma$, which clearly mirrors what was employed to map the monocoalesced periodic case onto the isolated-crest problem. Hence (\ref{eq:7.10}) rests on the celebrated composition
law for hyperbolic tangents ($\tau_1,\tau_2$): $\tau_1\ast\tau_2 = (\tau_1+\tau_2)/(1+\tau_1
\tau_2)$. It would be interesting to know whether the associated group properties give access to
still more general solutions to the Sivashinsky equation (\ref{eq:siveq}). That the scale-invariant signum function featured
in
(\ref{eq:2.5}) (\ref{eq:2.6}) is left unchanged by the successive changes of variables also is a key
property that traces back to the presence of the Hilbert transform $\hat{H}(-\phi_x)$ in
(\ref{eq:siveq}): {\it in fine}, it expresses that the complex velocity about the flame is a
sectionally-analytic function in the complex $x$-plane, which is indeed  a robust statement for it
is little affected by conformal changes of variables that would leave the real axis globally
invariant.

Normalizing $P(B)$ to $2N$ brings about the grouping $N\nu$ and, as long
as the integral equations (\ref{eq:2.5}) (\ref{eq:2.6}) of the continuous approximation are concerned, there is no reason why $N$ should be an integer.
Thus, (\ref{eq:2.5}) (\ref{eq:2.6}) effectively admit a 1-parameter continuum of solutions. The
situation is -- in a sense, analogous to the Saffman-Taylor problem of viscous fingering and
related
ones (see \cite{pelceII} and the references therein): when surface effects (here curvature) are
omitted, a
continuum of steady patterns is found. The equation (\ref{eq:siveq}) for flames is peculiar, however, because one
knows from the very beginning that only a discrete set of steady monocoalesced solutions exist,
corresponding to $N$s that are integers less than a well-defined $\nu$-dependent value, $N_{\mathrm{opt}}(\nu)$.
The Sivashinsky equation (\ref{eq:siveq}) thus offers the opportunity to see how the WKB
approaches to finger-width selection developed \cite{pelceII}
for the Saffman-Taylor problem, or kin, can be transposed to the present system to obtain a
quantization
condition on $N\nu$; for here ``inviscid'' solutions are now available and one knows the answer in
advance. This analysis likely is a key step to study flame response to noise, but has not yet been
completed. Because WKB approaches essentially look for solutions of a linearised equation in the
form $\exp(i \int^x k(x')\,dx')$, where $k(x)=O(1/\nu)$ depends on the ``inviscid'' solution, it is
seen that obtaining the latter to leading ($O(1)$) order in $\nu$ is not enough. Hence the
question: how to compute the leading ($O(\nu)$?) correction to the flame profiles obtained above?
Obviously this would require to better understand the nature of the continuous approximation
leading to the integral equations (\ref{eq:2.5})-(\ref{eq:2.7}) or (\ref{eq:7.4a}) (\ref{eq:7.4b}) for pole densities. In this context one may
perhaps adopt the -- rather unusual -- view point that the exact pole equations (\ref{eq:2.4}), once
specialized to $z_{\alpha} = i B_{\alpha}$ and steady patterns, constitute Gauss-like 
quadrature formulae to evaluate (\ref{eq:2.5}) numerically. How to define a ``best'' way of choosing
the pivotal values, {\it i.e.} the $B_{\alpha}$s, naturally leads \cite{specialfunctions} to the
notion of orthogonal polynomials associated with the Sivashinsky equation (\ref{eq:siveq}). In the case of Wigner's equation
(\ref{eq:3.2}) the Hermite polynomials are invoked \cite{mehta}, but we are not aware of such
mathematical analyses about (\ref{eq:siveq}) and (\ref{eq:2.5}) (\ref{eq:2.6}).

Next, we recall that 2-crested patterns studied in Sec.VII also belong to a continuous family of
solution profiles, now indexed by two independent parameters $N\nu, n\nu$. Even if $N+n$ is assumed
to be given by the optimum value $N_{\mathrm{opt}}(\nu)\simeq 1/2\nu$, there still remains the
question of how $N/n$ is selected in numerical resolutions of (\ref{eq:siveq}) with Neumann
conditions at $x=0$, $\pi$. The ratio $N/n$ can undoubtedly be chosen by the initial flame shape
$\phi(x,0)$. In the
case of forced propagations, the noise intensity ($\mu$) might well play also a role, for one can
imagine situations where $\exp(-\pi/2\nu) \ll \mu \ll \exp(-\pi/4\nu)$: the noise is then
intense
enough to break monocoalesced patterns (see {\it Introduction}), yet too weak to noticeably affect
the more curved 2-crested patterns with $N=n$.

To tailor a global criterion as to compare the 2-crest patterns and their response to noise, the
following remarks could be of some use. Let us collectively denote the $B_{\alpha}$s and
$b_{\gamma}$s as $\boldsymbol{B}$ and $\boldsymbol{b}$, respectively. The unsteady versions of
(\ref{eq:7.3a}) (\ref{eq:7.3b}) -- the pole equations for bicoalesced patterns -- may be re-written as 
\begin{equation}
	\label{eq:con1}
	\displaystyle
	\frac{d\boldsymbol{B}}{dt'} = -\boldsymbol{\nabla}_{\boldsymbol{B}} U, \qquad
\frac{d\boldsymbol{b}}{dt'} = -\boldsymbol{\nabla}_{\boldsymbol{b}} U,
\end{equation}
in terms of $U(\boldsymbol{B},\boldsymbol{b}) = V(\boldsymbol{B}) + v(\boldsymbol{b}) +
w(\boldsymbol{B},\boldsymbol{b})$, with
\begin{eqnarray}
	V(\boldsymbol{B}) 		& = &
\nu \sum_{\alpha}\left|B_{\alpha}\right|-2\nu^2\sum_{\alpha,\,\beta <
\alpha}\log\left|\sinh\frac{B_{\alpha}-B_{\beta}}{2}\right|,\label{eq:con2}\\
	w(\boldsymbol{B},\boldsymbol{b}) & = &
-2\nu^2\sum_{\gamma,\,\beta}\log\left(\cosh\frac{b_{\gamma}-B_{\beta}}{2}\right),\label{eq:con3}
\end{eqnarray}
and an expression similar to (\ref{eq:con2}) for $v(\boldsymbol{b})$; $t'=t/\nu$ is time scaled by
the shortest growth time of small-scale wrinkles (see {\it Introduction}). Accordingly, when the
right-hand sides of (\ref{eq:con1}) are supplemented with statistically identical, independent
random ({\it e.g.}, Gaussian) additive forcings, the joint probability density of
$(\boldsymbol{B},\boldsymbol{b})$ will tend to a quantity $\sim
\exp(-U(\boldsymbol{B},\boldsymbol{b})/\mu^2)$, where $\mu \ll 1$ characterizes the noise
intensities.
Because $U \sim 1$ in the small-$\nu$ limit (since $P(B)$ and $p(b)$ are $O(1/\nu)$) the above exponential is strongly
peaked
about the steady solutions. One can thus think of employing the $N/n$-dependent scalar
$U(\boldsymbol{B},\boldsymbol{b})$, evaluated at steady state, as an objective means to
discriminate the various 2-crested patterns in the presence of forcing. The task of evaluating
$U$ in the continuous approximation has not yet been completed. Neither is the analysis required
to handle situations where the poles are slightly misaligned... yet still symmetric about $x=0$ and
$x=\pi$ for compatibility with Neumann boundary conditions.

One must finally stress that the present analyses did not exhaust all the possibilities of
``steady'' solutions of (\ref{eq:siveq}), even with $2\pi$ as minimal periodicity. The {\it
interpolating solutions} discovered in \cite{denet06,guidi03} constitute another class, comprising
(possibly many-) extra poles, nearly evenly distributed \cite{vm,procaccia99} along sinuous curves
at a distance from the real axis. In our opinion such unstable equilibriums are also worth analyzing
in detail for $\nu\rightarrow 0$, as are those mentioned in Sec.\ref{sec:dynamics} and generalizations of (\ref{eq:siveq}) itself \cite{joulin91}.

As an end to a numerical work on (\ref{eq:siveq}), with noise included in the right hand side
\cite{denet06}, one of us concluded that ``\ldots it is likely that new analytical studies of the
Sivashinsky equation should be possible: even if the equation is now almost 30 years old, many
things remain to be explained". The words still hold true.
\begin{acknowledgments}
One of us (G.J.) thanks H. El-Rabii (Poitiers University) for discussions, helps in the Calculus...
and the \LaTeX~	version of the paper.
\end{acknowledgments}


~\newpage

\noindent{\large\bf List of figures}\\

\begin{figure}
 \includegraphics[width=\textwidth]{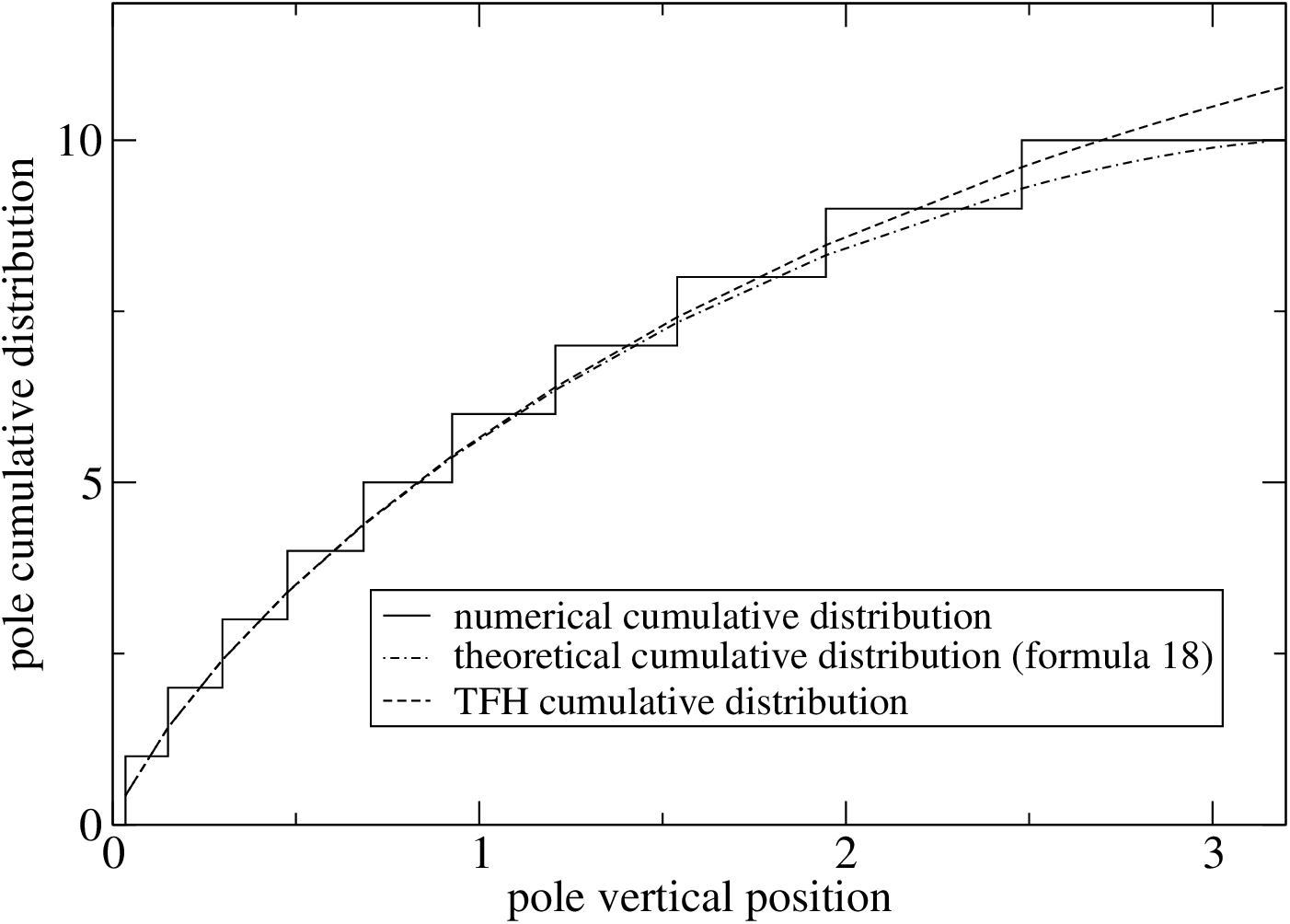}
 \caption{\label{fig:1} Numerical vs analytical cumulative pole densities, for an isolated crest
with $1/\nu = 19.5$, $N = 10$. If exact, the theoretical curve (dot-dashed line, eq.
(\ref{eq:3.7}))
would pass through the middle of the risers of the numerical staircase (solid line, eq.
(\ref{eq:2.4})). The dashed line is the TFH fit.}
\end{figure}

\begin{figure}
  \includegraphics[width=\textwidth]{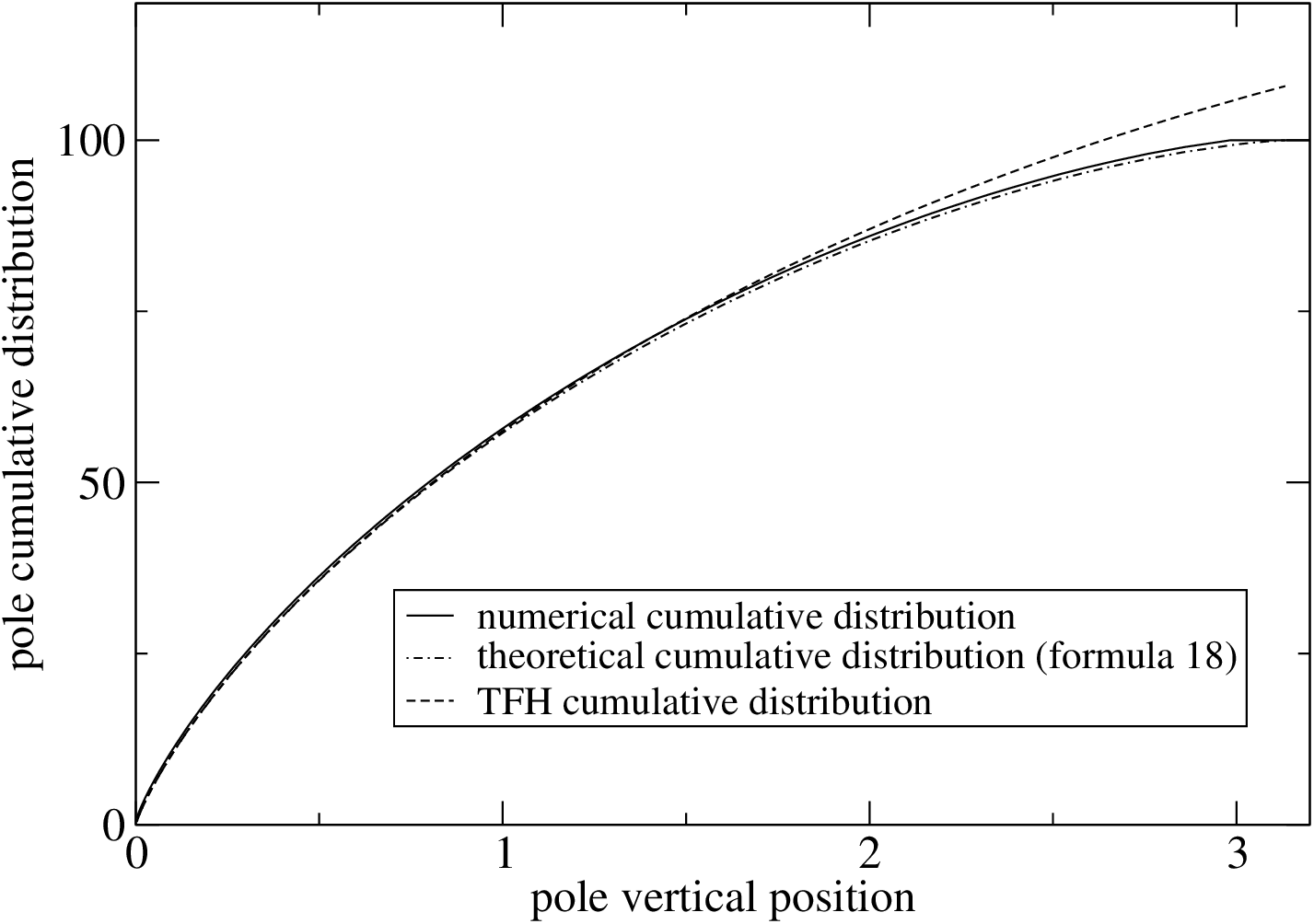}
 \caption{\label{fig:2}Same as in Fig. \ref{fig:1}, for $1/\nu = 199.5$, $N = 100$. Only the
upper hull (solid line) of the exact staircase is shown, for readability. The dashed line is the
TFH fit.}
\end{figure}

\begin{figure}
  \includegraphics[width=\textwidth]{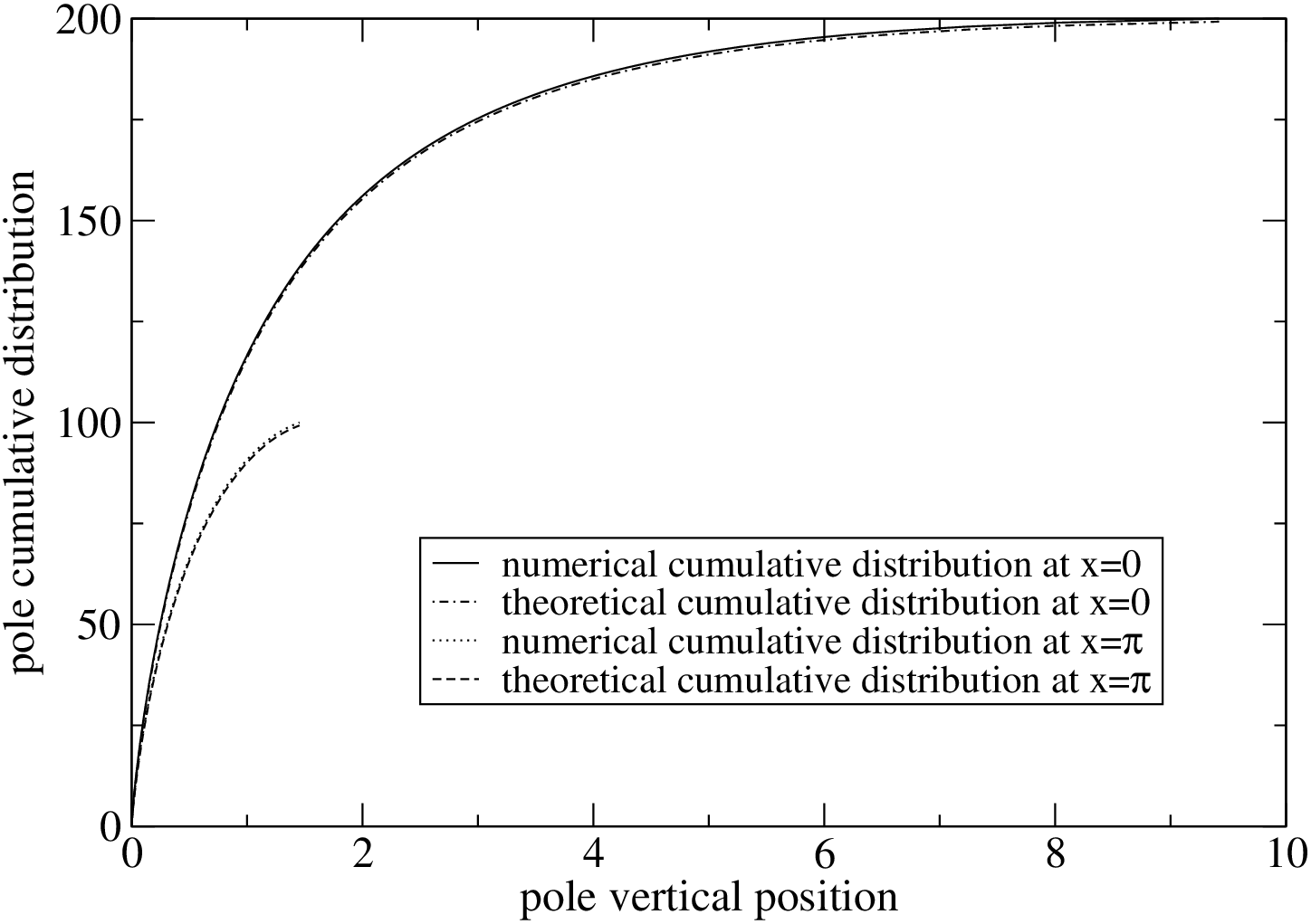}
  \caption{\label{fig:7}Cumulative pole densities $R(B)$ (upper curves) and $r(b)$ for a
bicoalesced
periodic pattern with $1/\nu = 600.5$, $N = 200$, $n = 100$: the solid and the dotted lines are
from
eqs. (\ref{eq:7.7a}) (\ref{eq:7.7b}) and (\ref{eq:7.9}) (\ref{eq:7.10}); the dashed and the
dot-dashed ones are the upper hulls of the exact staircases (see Fig. \ref{fig:1}). As
$(N+n)=N_{\mathrm{opt}}(\nu),$ $B_{200}=\infty.$}
\end{figure}

\noindent{\bf Fig.\phantom{.}1:} Numerical vs analytical cumulative pole densities, for an isolated
crest
with $1/\nu = 19.5$, $N = 10$. If exact, the theoretical curve (dot-dashed line, eq. (\ref{eq:3.7}))
would pass through the middle of the risers of the numerical staircase (solid line, eq.
(\ref{eq:2.4})). The dashed line is the TFH fit.\dotfill 23\\\\
{\bf Fig.\phantom{.}2:} Same as in Fig. \ref{fig:1}, for $1/\nu = 199.5$, $N = 100$. Only the
upper hull (solid line) of the exact staircase is shown, for readability. The dashed line is the
TFH fit.\dotfill 24\\\\
{\bf Fig.\phantom{.}3:} Numerical ((\ref{eq:3.8b}), solid line) vs analytical ((\ref{eq:3.6d}),
dot-dashed line) pole densities $P(B)$ for an isolated crest with $1/\nu = 199.5$, $N = 100$. The
dashed line is the TFH fit.~ \dotfill 25\\\\
{\bf Fig.\phantom{.}4:} Shapes of an isolated crest with $1/\nu =19.5$, $N = 10$: continuous
approximation ((\ref{eq:25}), dot-dashed line), exact (solid line), and smooth approximation from
eq. (\ref{eq:3.10}) (dotted).\dotfill 26\\\\
{\bf Fig.\phantom{.}5:} Numerical (solid line) vs analytical (eq. (\ref{eq:4.4}), dot-dashed line)
cumulative pole densities for a monocoalesced periodic crest, for $1/\nu = 199.5$, and $N = 100$
($=N_{\mathrm{opt}}(\nu)$). Only the upper hull of the numerical staircase is shown.\dotfill 27\\\\
{\bf Fig.\phantom{.}6:} Shapes of a monocoalesced periodic flame with $1/\nu = 19.5$, $N = 10$
($=N_{\mathrm{opt}}(\nu)$): continuous approximation ((\ref{eq:4.11}), dot-dashed curve) vs exact
result (solid line) and smooth approximation ((\ref{eq:4.8}), dotted).\dotfill 28\\\\
{\bf Fig.\phantom{.}7:} Cumulative pole densities $R(B)$ (upper curves) and $r(b)$ for a bicoalesced
periodic pattern with $1/\nu = 600.5$, $N = 200$, $n = 100$: the solid and the dotted lines are from
eqs. (\ref{eq:7.7a}) (\ref{eq:7.7b}) and (\ref{eq:7.9}) (\ref{eq:7.10}); the dashed and the
dot-dashed ones are the upper hulls of the exact staircases (see Fig. \ref{fig:1}). As
$(N+n)=N_{\mathrm{opt}}(\nu),$ $B_{200}=\infty.$\dotfill 29\\\\
{\bf Fig.\phantom{.}8:} Theoretical pole densities $P(B)$ (resp. $p(b)$) plotted as dot-dashed 
or dashed lines vs $\theta$, eq. (\ref{eq:7.9}), with $\sin\sigma$ replaced by $(\tanh B/2)/A$
(resp. $(\tanh b/2)/a$). The solid and the dotted lines are the numerical pole densities. All are
for a bicoalesced periodic flame with $N=200$, $n=100$, $1/\nu=600.5.$\dotfill 30\\\\
{\bf Fig.\phantom{.}9:} Shapes of a bicoalesced flame with $1/\nu = 199.5$, $N = 80$, $n = 20$:
exact (solid line) vs continuous approximation (from integration of (\ref{eq:7.15a}),
dot-dashed).\dotfill 31\newpage

~\newpage

~\newpage

~\newpage

~\newpage

~\newpage

~\newpage

~\newpage

~\newpage

\end{document}